\documentstyle[12pt,epsf]{article}

\def\wsh{{\Sigma_g}}                                   
\def\tsp{{\cal M}}                                     
\def\Dcal{{\cal D}}                                    
\def\Zm{{{\cal Z}_\tsp}}                               
\def\lmst{{\lambda_{st}}}                              

\def\gab{{g_{\alpha\beta}}}                            
\def\Guv{{G_{\mu\nu}}}                                 

\def\pa{{\partial_\alpha}}                             
\def\pb{{\partial_\beta}}                              
\def\Xu{{X^\mu}}                                       
\def\Xv{{X^\nu}}                                       
\def\xia{{\xi^\alpha}}                                 

\def\pXu{{\partial X^\mu}}                             
\def\pXv{{\partial X^\nu}}                             
\def\pxia{{\partial\xi^\alpha}}                        

\def\Udg{{U^\dagger}}                                  

\def\opfd{{{\cal O}_F}}                                
\def\opslr{{{\cal O}_P}}                               

\def\bra{\langle}
\def\ket{\rangle}

\def\hlf{{1 \over 2}}

\newcommand\binom[2]{{{#1}\choose {#2}}}

\newcommand\putfig[3]{
   \vbox{
   \let\picnaturalsize=N
   \def\picsize{#3}
   \def\picfilename{#1}
   \ifx\nopictures Y\else{\ifx\epsfloaded Y\else\input epsf \fi
   \let\epsfloaded=Y
   \centerline{\ifx\picnaturalsize N\epsfxsize \picsize\fi
   \epsfbox{\picfilename}}}\fi
   \vspace{1.0cm}
   {\it #2}
   \vspace{1.5cm}
   }
}

\def\cmp#1{{\it Comm. Math. Phys.} {\bf #1}}

\def\pl#1{{\it Phys. Lett.} {\bf #1B}}
\def\prl#1{{\it Phys. Rev. Lett.} {\bf #1}}
\def\prd#1{{\it Phys. Rev.} {\bf D#1}}

\def\np#1{{\it Nucl. Phys.} {\bf B#1}}

\def\jmath#1{{\it J. Math. Phys.} {\bf #1}}

\begin{document}

\begin{titlepage}
\titlepage
\rightline{TAUP- 2152-94}
\rightline{\today}
\vskip 1cm
\centerline {{\Large \bf Folds in 2D String Theories}}
\vskip 1cm

\centerline {O. Ganor , J. Sonnenschein and S. Yankielowicz
\footnote{Work supported in part by the US-Israel Binational Science
Foundation and the Israel Academy of Sciences.}}

\vskip 1cm

\begin{center}

\em  School of Physics and Astronomy\\

Beverly and Raymond Sacler \\

Department of Exact Sciences\\

Tel-Aviv University\\

Ramat Aviv Tel-Aviv, 69987, Israel

\end{center}

\vskip 1cm

\abstract{
We study
 maps from a 2D world-sheet to a 2D target
space which include folds. The geometry of folds is discussed and a metric
on the space of folded maps is written down. We show that the latter is
  not invariant under
area preserving diffeomorphisms of the target space. The contribution
to the partition function of
 maps associated with a given fold configuration  is computed. We derive
a description of folds
in terms of Feynman diagrams. A scheme to sum up the contributions of folds
to the partition function in a special case is
suggested and is shown to be related to  the Baxter-Wu
  lattice model.
An interpretation  of folds as trajectories of particles in
 the adjoint representation of  $SU(N)$  gauge group in the large $N$ limit
which interact in an  unusual way with  the gauge fields
 is discussed.}

\end{titlepage}

\section{Introduction}
We study the space of  maps, including folds,
of a two dimensional  Nambu-Goto (NG) string theory.
The recent  renewed interest in the NG theory was triggered
by the  discovery of stringy behavior of $SU(N)$
 Yang-Mills theory  in the large $N$ limit \cite{Gross,GT1,GT2}.
The latter theory  was shown to be equivalent to
a 2D string theory. This  equivalence is based on
identifying the  string coupling constant  with ${1\over N}$ and the string
tension with ${1\over 2}\tilde g^2 N$ where $\tilde g$ is the gauge coupling.
The asymptotic expansion of the $YM_2$ partition function was recast in terms
of a sum of maps from the compact Riemann surface world-sheet to a  2D target
space \cite{Gross,GT1,GT2}.
 Point singularities of these maps, like branch points played an
important role in the correspondence between the 2D string theory and the
$YM_2$ theory. However, line singularities,  which will be
shown to correspond to folds, had to be
excluded by hand. It is the absence of folds which accounted for the fact
that pure 2D $YM_2$ theory does not include propagating particles.
Using the same logic one is led to consider folds while searching for the
stringy picture of $QCD_2$.
Indeed an  analysis  of folded maps  both classically and in a
framework of canonical quantization revealed a behavior of  particle
trajectories \cite{Bars}.
Following the work of D. Gross and W. Taylor\cite{Gross,GT1,GT2},
an intensive research effort was devoted to the stringy behavior of two
dimensional YM theory \cite{Mina,SYM2}

 A natural candidate for the stringy YM action  would have been the Nambu Goto
action. However,  for a two dimensional target space the NG action is
equivalent to a sum of a topological number,
the winding number, and a term that counts the area
enclosed by folds. Hence, the NG theory  cannot be associated, without
further alteration, with stringy $YM_2$.
Up to date the only consistent method to quantize
the NG theory was to translate
the classically equivalent formulation of Polyakov\cite{Poly}
 and quantize the latter. For a two
dimensional target space this is the Liouville continuum
formulation of the non-critical string theory.
As was discovered in the work of Gross and Taylor\cite{GT1,GT2}
singular maps play a major role in performing the sum of maps that corresponds
to the partition function of the $YM_2$ theory.
The action of Polyakov\cite{Poly} is not an adequate scheme to discuss these
types of maps. In fact, it is easy to realize that gauge fixing the world-sheet
reparametrization in the conformal gauge (or the world-sheet light-cone gauge)
forbids altogether folds and certain point singularities.

Another path which was followed in the search for a first quantized
action of stringy $YM_2$ was that of topological string models.
A positive sign in this direction was the
fact that the torus  YM chiral partition function was reproduces in terms  of
a summation over holomorphic maps from a toroidal worldsheet\cite{BCOV}. Indeed
proposals for a topological sigma model action of stringy YM theory were
recently written down\cite{Horava,Moore}.

 Our motivation to study folded maps was thus two folded:
(i) Folds as a possible framework to describe particle trajectories
in the context of stringy  $QCD_2$.
(ii) The contribution of folded maps
to the functional integral quantization  of  the NG theory.

In the present work we study
 maps from a compact  Riemann surface to a compact Riemann surface, the
 world-sheet and  the target space respectively.
The  main  idea  is to explore the space of maps including  folded maps.
 Unfolded maps   are characterized by the numbers of sheets with the same
and opposite
orientations as that of the
target space, and  by their singular points. The latter
include branch points, tubes that connect sheets of the cover with
the same and opposite orientations, and contracted handles\cite{GT2}.
 Folds are  boundaries of ``smooth regions'' which are covered with sheets of
both orientabilities.  Along these boundaries,
the sheets of two regions are continued with a (possibly)
non-trivial permutation among the sheets, and also,
two other sheets of opposite
orientability and of the same side of the fold are joined by the fold.
Folds are thus characterized by their trajectories on the target space, and by
certain types of  singular points on them. An algebraic as well as
pictorial  description of the latter is presented in our work.

An essential step in the analysis of folds  is to write down the
 full measure of  the  generating functional
which includes folds. The measure  is induced from the metric
on the space of maps  that
is itself naturally induced from the maps.
We found out that the metric  on the space of maps includes in addition to
the metric of unfolded maps two additional terms,
one which is expressed in terms
of an integral over the proper length of the fold,
and another one which involves the singular points on the fold.
Due to these two additional terms  the corresponding measure  ceases to be
invariant under  area preserving diffeomorphisms of the target space.
Note that once we incorporate folds the correspondence to a $YM_2$ theory is
lost. Thus, non-invariance does not create any difficulty. Moreover, if folds
are associated with particles interacting with the gauge fields, there is no
reason to anticipate such an invariance.
It is emphasized in our work  that the measure of the full NG theory
has many undetermined parameters
and  thus it defines in fact a large family  of string models.

The nature of folds as particle trajectories is demonstrated in the computation
of the contribution to the partition function of  maps from a sphere to a
sphere with winding number one and a single fold. This result is then
generalized to the case where the bounded region is covered by an arbitrary
number of covers from both orientabilities.
We develop  a set of Feynman
diagram rules for
maps that include several, possibly intersecting folds.
 These include expressions for fold propagators, vertices of
intersecting folds, the factors which are associated with smooth regions etc.
The implementation of these Feynman diagram method to determine the partition
function seems to be a difficult task.
The expansion parameter is discussed.
For the case of a fold enclosing one
cover of each orientability, we translate the problem to that of an Ising like
model perturbed by triple interaction terms.
The interpretation of folds as particles was further clarified by applying
the techniques developed in \cite{GT2} to handle Wilson lines. Using the
``generalized Frobenious characters" we found that the operator which is
associated
with a fold could be understood as  a  trajectory of a particle in
 the adjoint representation with an unusual  interaction with the gauge
fields.

The paper is organized as follows.
In section
(2) we present the target space picture of the $2D$ string theory
via the NG formulation. Folds and point singularities of these  maps
are described algebraically as well as pictorially. In particular we analyze
singular points on folds, branch points and ``quartic points".
 Section (3) is devoted to the measure on the space of maps.
We start with the measure of unfolded maps,
emphasizing that there are many undetermined parameters
 in the model and hence it   defines in fact a large family of string theories.
We then write down a metric on the space of folded maps and show that it
introduces non-invariance under target space area preserving diffeomorphisms.
World sheet reparametrization is discussed. We show that standard gauges are
not adequate.

In section (4)
a computation of the contribution of a fold to the  partition function is
presented.  This is done at first for the simplest fold and then generalized
to the case of an arbitrary number of covers
via the introduction and
 calculation of  the ``fold transition matrix''. We discuss the
intersection of folds, introduce
the notion of ``quartic points" and write down a set of
Feynman  diagrams.  The summation over fold configurations is the  topic
of section (5). We
introduce a scheme for performing this sum
 which is related, at least for the case of low
cover numbers,  to an Ising-like model with triple interactions
(known as the Baxter-Wu model)\cite{BAXWU}.
 Section (6) is
devoted to the  interpretation of the folds as particle trajectories and the
correspondence to  the string description of $YM_2$ theory\cite{GT1,GT2}.
We use the tools developed for
evaluating  Wilson lines to handle folds.
It is shown that folds behave like trajectories of
particles in the adjoint representation of a $SU(N)$ gauge group in the large
$N$ limit, and that  singular
points on folds introduce peculiar interaction with the jump in $Tr(F^2)$
along the fold.
In section (7) we summarize, present some conjectures and state certain open
questions.
Explicit calculations of the fold matrix are presented in the appendices. In
Appendix A  the fold matrix for a fold with no branch points is computed.
Appendix B is devoted to the calculation of the eigenvalues of the
``fold-matrix ''and the ``restricted fold matrix''.

\section{The target space picture}
Let $\xia$, $\alpha=1,2$ denote the coordinates on $\wsh$, a genus $g$
world sheet, and $X^\mu(\xia)$ denote a map from $\wsh$ to $\tsp$, a $D=2$
dimensional target space.
The most natural  framework  to   study these maps
is the Nambu-Goto (NG) string theory  for which the partition function is given
by \begin{equation}
\Zm = \sum_g \lmst^{2g-2}\int\Dcal X e^{\int d^2\xi\sqrt{\det(\gab)}}
\end{equation}
where
\begin{equation}
\gab=\Guv(X)\pa\Xu\pb\Xv     \label{GAB}
\end{equation}

is the metric on the world sheet induced from the target space metric,
$\Guv(X)$ is the target space metric and $\lmst$ is the string coupling
constant.
In fact for target spaces of dimension $D=2$ the  NG action takes
a simpler form

\begin{equation}
S=\int d^2\xi\sqrt{\det(\Guv)}\, |\det({\pXu\over\pxia})|.
\label{mishac}
\end{equation}

Classically the NG action is equivalent to the action written in the
formulation of Polyakov \cite{Poly}.
So far the only consistent way to quantize the continuum  2D
 string theory has been using the latter approach.
As will be shown later, Polyakov's string is not an adequate arena to analyze
folds. It is thus
the idea of the present work  to treat the quantum theory  in the NG framework.

It is well known that the action and measure are invariant under
area preserving diffeomorphisms (APD).
The invariance of the measure $\Dcal X$ under
$X(\xi)\rightarrow Y(X(\xi))$ where $Y(X)$ is a coordinate transformation
on the target space follows because the change of the measure obeys
$$
{{\Dcal X}\over{\Dcal Y}} =
\prod_\xi\det({{\partial Y^\mu}\over\pXv})
$$
where for APD the determinants are 1.
The action (\ref{mishac}) is obviously
also unmodified by these diffeomorphisms.
However, as will be shown below,
 this does not mean that the NG theory itself has the APD symmetry.

According to whether the determinant  of the Jacobian matrix
\begin{equation}
J_\alpha^\mu={\pXu\over\pxia}
\label{mishJ}
\end{equation}
vanishes or not, we classify the maps $X^\mu(\xia)$ into two types.
Unfolded maps correspond to $det (J)\neq  0$ everywhere.
Folds, as will be clarified in
the following section, are curves on the target space on which $det (J) =0$.
It is only when we restrict to the space of unfolded maps
that the theory becomes invariant under this symmetry.
 In section (3.2) it is shown
that folds turn
the measure $\Dcal X$ into a  singular one  and in order to correct its
singularity we have to break APD invariance.


Let us proceed  with a short reminder on unfolded maps. These maps are taken
to be locally covering maps of $\tsp$ apart from a finite set of singular
points. They are characterized by the number of sheets with the same
orientation as that of $\tsp$ and by the number of sheets of opposite
orientation. In addition the maps are classified according to their singular
points. The singularities include branch points, contracted tubes that
connect sheets of the cover with the same and opposite orientations, and
contracted handles\cite{GT2}.
It was shown by D. Gross and W. Taylor\cite{GT1,GT2} that
the  $SU(N)$ $YM_2$  theory at $N\rightarrow\infty$ is reproduced by summing
over unfolded maps which include  branch points,
 contracted tubes, contracted handles as well as $2g-2$ ``$\Omega$-points''.

\subsection{Geometry of the Folded Maps}

Let us start with a pictorial description   of certain geometrical
structures that arise with folded maps.
The general form in target space of a map $X:\wsh\rightarrow \tsp$ is of
regions with no folds that are bounded by folds (Fig.1). The folds are thus
considered as boundaries of ``smooth regions''.
In each such region there might be a multiple  cover with sheets of both
orientabilities.
For instance in Fig.2, the region under the ``pocket'' has three sheets
one with a $(-)$ orientability and the other two with a $(+)$ orientability.
The region under the ``smashed handle'' of Fig.3 has the same cover-structure.
Along the folds, the sheets of two regions are continued with a (possibly)
non-trivial permutation among the sheets, and also, two sheets of opposite
orientability and of the same side of the fold are joined by the fold.
Thus, for each fold one side of it has two more sheets than the other --
one sheet more for each orientability.

\putfig{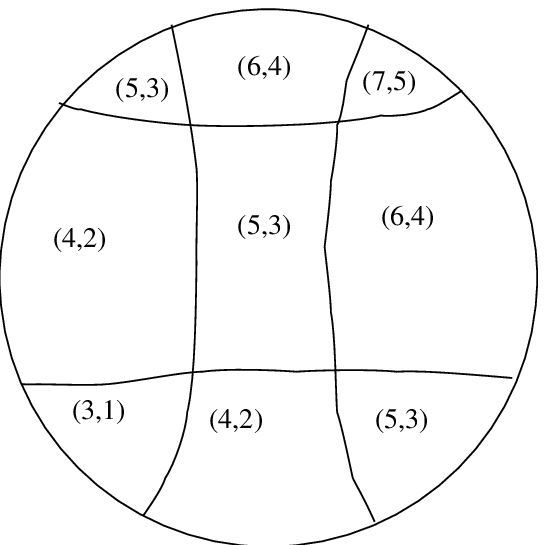}{ Fig. 1: A part of a target space that is covered
  by a map with folds. The areas \mbox{$(m,n)$} represent regions with
  \mbox{$m$} sheets of \mbox{$(+)$}
  orientation and \mbox{$n$} sheets of \mbox{$(-)$} orientation. }{50mm}

\putfig{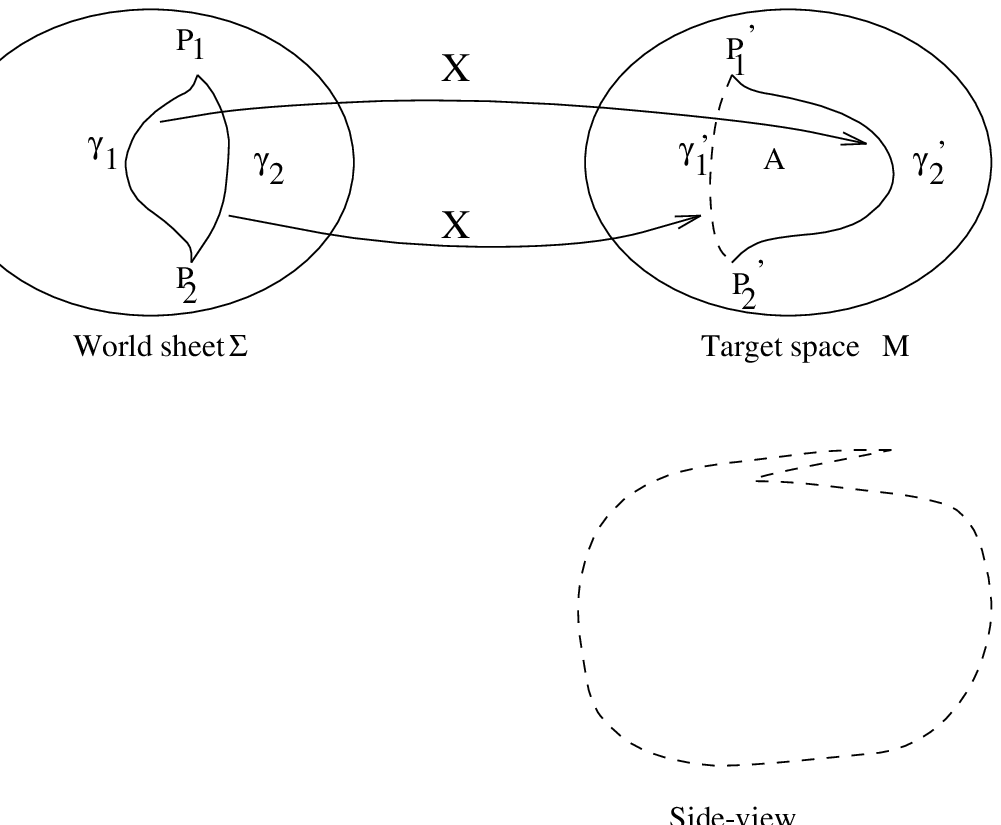}{ Fig. 2: A ``pocket'' is formed when an area $A$ in
  target space is a \mbox{$3$}-sheet cover
  inside a \mbox{$1$}-sheet cover \mbox{$M-A$}.
  The boundary curve has two
  singular points \mbox{$P_1,P_2$} where the joining
  of the sheets changes.}{90mm}

\putfig{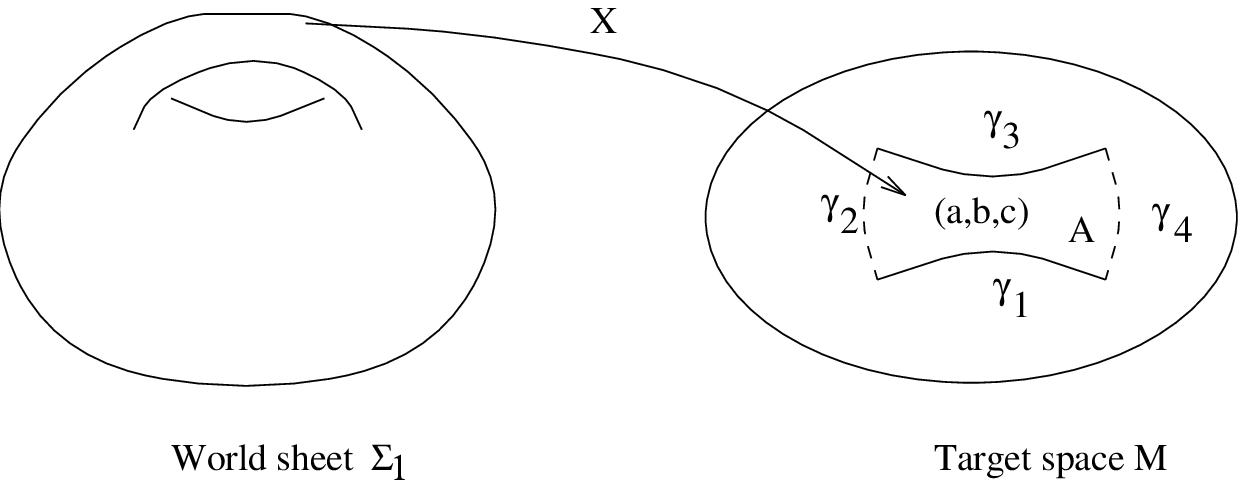}{Fig. 3: A ``smashed handle'' is formed when a
  \mbox{$3$}-sheet area \mbox{$A\subset M$} is bounded
  by a curve with \mbox{$4$} singular points.
  Along \mbox{$\gamma_1, \gamma_3$} -- \mbox{`$a$'}
  and \mbox{`$b$'} are joined and \mbox{`$c$'}
  continues smoothly to \mbox{$M-A$}.
  Along \mbox{$\gamma_2, \gamma_4$} -- \mbox{`$c$'} and \mbox{`$b$'}
  are joined and \mbox{`$a$'} continues smoothly to the outside.}{90mm}

A fold is a curve on which
$det(J^\mu_\alpha)=0$.
Along the folds there might be several singular points where the joining
of the sheets changes as can be seen in Fig.2.
It might seem that such points are singular also geometrically, that is
they are points at which the direction of the tangent along the fold is
discontinuous. It is shown in what follows that generically this is {\em not}
the case . Namely, we will see that (generically) points with a
discontinuous tangent are points where the Jacobian matrix
$J^\mu_\alpha$
vanishes altogether, whereas along folds only the determinant
$\det(J_\alpha^\mu)$ vanishes, and at singular points along a fold the
determinant has in some sense (to be made precise later on) a double zero.

Let us analyze algebraically the possible singular  points of the
general map
\begin{eqnarray}
X_1 &=& f(\xi_1,\xi_2) \nonumber\\
X_2 &=& g(\xi_1,\xi_2)
\end{eqnarray}
at the vicinity of the origin $\xi_1=\xi_2=0$.
Expanding in a Taylor series we get after performing a local world sheet
reparametrization the generic map
\begin{eqnarray}
X_1 &=& \xi_1 \nonumber\\
X_2 &=& \xi_2
\end{eqnarray}
which is a point inside a smooth region.

The other cases happen when some of the derivatives of $f$ or $g$ vanish.
In each case we  expand in leading order in $\xi_1,\xi_2$, performing
a world-sheet reparametrization and a target space Lorentz rotation  and
a parity transformation if necessary.
We will restrict ourselves to situations where up to two
conditions are imposed on the derivatives of $f$ and $g$ because a generic
map $\Xu(\xia)$ has no points on the world-sheet where three conditions
are satisfied (three equations in two variables).
In what follows $A,B,C,\dots$ are parameters.
The various possible cases  are:
\begin{enumerate}
\item {\em A simple point on a fold}

This is the case when $\det(J_\alpha^\mu) = 0$. We can bring
the map to a form
\begin{eqnarray}
X_1 &=& \xi_1^2 + A\xi_2 + \cdots \nonumber\\
X_2 &=& \xi_2
\end{eqnarray}
where $X_1$ has no linear term in $\xi_1$.
This map has a fold in the vicinity of $(0,0)$ on the curve $X_1=AX_2$.
It describes two sheets below it and no sheets above it.

\item {\em A singular point on a fold}

We can supplement the condition $\det(J_\alpha^\mu) = 0$ with a
further condition as follows. If $J_\alpha^\mu \neq 0$ as a matrix, then
there is a direction $a^\alpha\pa$ in the tangent space
to the world-sheet at the origin $T\wsh$ such that
$$
a^\alpha\pa\Xu = 0,\qquad \mu=1,2
$$
We demand that
$$
a^\beta\pb \det(J_\alpha^\mu) = 0
$$
as well.
For a generic point on a fold the direction $a^\alpha\pa$ is different
from the fold direction (the direction in which $\det(J_\alpha^\mu)$ remains
zero), but for the singular point that we discuss now the direction
$a^\alpha\pa$ coincides with the direction of the fold.
In the vicinity of such a point the map looks like
\begin{eqnarray}
X_1 &=& \xi_1^3 - A\xi_1\xi_2 + B\xi_1^2\xi_2 + C\xi_2 + D\xi_2^2 + \cdots
\nonumber\\
X_2 &=& \xi_2
\label{SINGMAP}
\end{eqnarray}
The qualitative analysis is not affected by taking $B=C=D=0$ for simplicity.
In that case the equation of the fold is
$$
3\xi_1^2 -A\xi_2=0
$$
The direction $a^\alpha\pa$ is $\partial_1$ which is also the direction of
the fold (in the world sheet).

In target space the fold looks like
$$
X_1^2 = 4({A X_2\over 3})^3 +O(X_2^{7\over 2})
$$
and we see that it has a smooth tangent vector (although it has an
infinite curvature at the origin).

Fixing $\xi_2$ and varying $\xi_1$ we obtain $X_1$ as a cubic function
of $\xi_1$. For $\xi_2 < 0$ the function is monotonous, and thus there is
one sheet in the target space below the curve $X_1^2 = {4\over 3}A^3 X_2^3$.
For $\xi_2 > 0$ the cubic function is not monotonous any more and passes
through each point above the curve $X_1^2 = 4({A X_2\over 3})^3$ thrice.
(See Fig.4).

\putfig{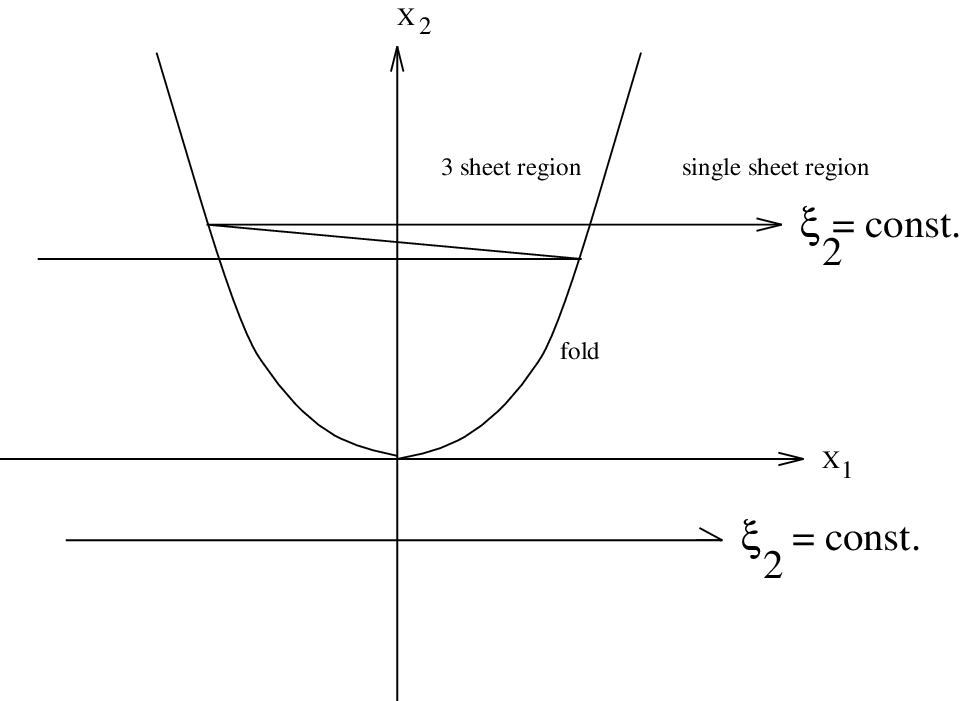}{Fig. 4: The map of equation (\ref{SINGMAP})
  for \mbox{$B=C=D=0$} creates a fold with a singular point at the origin.
  (The \mbox{$\xi_2 =const.$} line was split for the sake of visibility.
   It should have been overlapping.)}{80mm}
\bigskip
\item {\em A branch point}

When the Jacobian matrix vanishes two cases can arise.
They both correspond to maps of the form
\begin{eqnarray}
X_1 &=& A\xi_1^2 + 2B\xi_1\xi_2 + C\xi_2^2\nonumber\\
X_2 &=& D\xi_1^2 + 2E\xi_1\xi_2 + F\xi_2^2
\end{eqnarray}
substituting
\begin{eqnarray}
\xi_1 = r\cos t \nonumber\\
\xi_2 = r\sin t \nonumber
\end{eqnarray}
for an infinitesimal $r$ we see that a circle around $(0,0)$ in the world
sheet goes to an ellipse of winding number $2$ on the target space.
When the origin $(0,0)$ is inside the ellipse, $(0,0)$ is a branch point.
This happens when
$$
\Delta = A^2F^2+D^2C^2-4BEDC-4BEAF+4E^2AC+4B^2DF-2ACDF < 0
\label{BPCOND}
$$
Such a branch point is depicted in  Fig.5.

\putfig{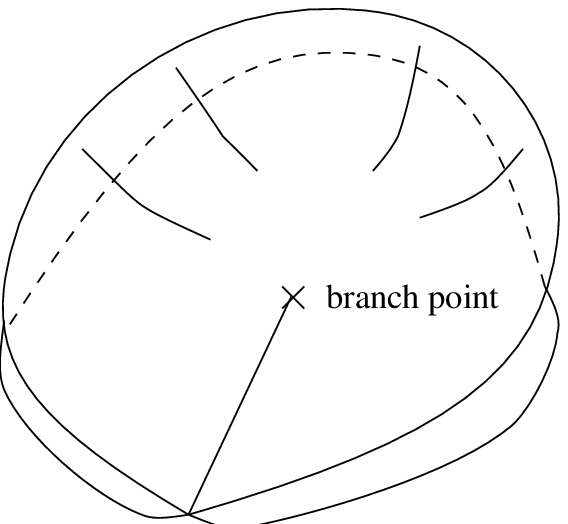}{Fig. 5: A branch point}{80mm}

\item {\em A quartic point}

The other case  corresponds to a map of the form
\begin{eqnarray}
X_1 &=& A\xi_1^2 + 2B\xi_1\xi_2 + C\xi_2^2\nonumber\\
X_2 &=& D\xi_1^2 + 2E\xi_1\xi_2 + F\xi_2^2\label{QUARTP}
\end{eqnarray}
with
$$
\Delta = A^2F^2+D^2C^2-4BEDC-4BEAF+4E^2AC+4B^2DF-2ACDF > 0
\label{QPCOND}
$$
We see that it corresponds to two double folds that meet at an angle.
They separate a region of four sheets from a region of no sheets.
Adding cubic terms to the functions $\Xu(\xia)$ we obtain a splitting
of the double folds and get four folds that meet at a ``quartic'' point
(See Fig.6).

\putfig{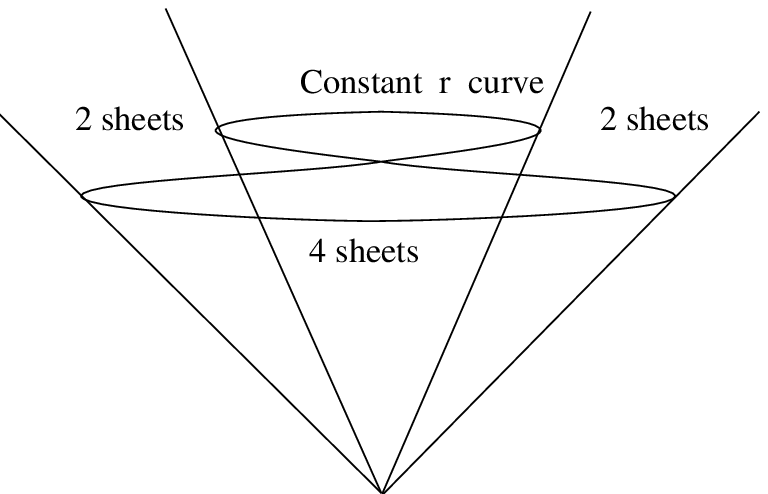}{Fig. 6: A quartic singular point (schematically).
  (The \mbox{$r =const.$} curve was perturbed
   for the sake of clearer visibility.
   It should have been overlapping.)}{50mm}

\end{enumerate}

The preceding discussion can obviously be generalized to maps which are
 higher order polynomials. This will introduce higher order  branch points.
Nevertheless, as was mentioned above, classically, generic maps do not have
those higher order branch points.

\section{The  Measure on the  space of Maps}
With the definitions of folds and singular points on them at hand, our next
object of interest is the measure on the entire
space of maps $\Xu(\xia)$ from
$\wsh$ to $\tsp$.
 It will be shown that
in order to incorporate the
geometrical configurations such as the  location of the folds and singular
points one cannot use the metric of the subspace of unfolded maps. Instead
one has to add to the latter two additional terms.
We start our discussion with the measure of the unfolded maps, introduce the
necessary modifications to incorporate folds and the singular points on them,
and analyze the invariance under APD.

\subsection{The Measure on the Space of Unfolded Maps}

Unfolded maps can be classified according to
 their  world-sheet independent properties namely,
the winding number and the location of branch-points.
On the other hand, the measure in the Nambu-Goto formalism
is proportional to $\prod_\xi dX(\xi)$
modulo world-sheet diffeomorphisms.
It is specified by specifying the corresponding metric
\begin{equation}
\| \delta X \|^2 =
\int \Guv\delta \Xu\delta\Xv\sqrt{\det(\gab)}d^2\xi
\label{METRDX}
\end{equation}
where $\gab$ is given in (\ref{GAB}).
The contributions of the singular points of unfolded maps to the measure
take the form of $\lambda dA$ where $dA$ is an infinitesimal target space
area in which the singular point is located and the $\lambda$-s are
parameters of dimension $1/A$ that specify the type of the singular point,
namely $\lambda_{br}$ for a branch point and (for the case of the string
theory of \cite{GT1}\cite{GT2})
$\lambda_{ct+},\lambda_{ct-}$ and $\lambda_{ch}$ for connecting
tubes of the same and opposite orientation and collapsed handles respectively.
This is not a consequence of (\ref{METRDX}) but has to be added ``by hand'',
as will be discussed in more detail in the next subsection.

Apriori we do not see any reason why those parameters have to be equal. The
latter case was found to be necessary for a string theory which corresponds to
the large $N$ limit of the $SU(N)$ $QCD_2$ theory. It seems to us that the
$\lambda$ parameters cannot be determined from the NG measure.

\subsection{ The Induced Measure on Folds}
What happens when we perturb a map $X(\xi)$ with no folds
with a small perturbation $\delta X(\xi)$ so that the map
$X+\delta X$ has an infinitesimal fold?
Although $\delta X$ is small, the derivative
$\nabla\delta X$ must be non-infinitesimal in order to create a fold
since we can reparametrize $\xi$ so that $\nabla X$ will be no-where
zero. Although the measure $\| \delta X \|$ can be small, we will
refer to maps with folds and maps with no folds as belonging
to different ``sectors''.

The measure on the space of maps,
 which is specified by the metric (\ref{METRDX}),
 is singular at maps with folds.
This happens because the metric (\ref{METRDX}) is not strictly positive
definite at maps with folds. At folds $\det(\gab)=0$ and thus
$\delta X(\xi)$ for $\xi$ on a fold is a zero mode of the metric.
Thus, the measure would be identically zero for maps  with folds.

Geometrically, a  folded map is specified by the location of its folds and
singular points. Thus it seems natural that the induced measure on maps modulo
world-sheet diffeomorphisms will be the induced measure on
the subset of infinitesimal changes
$\delta X(\xi)$ for $\xi$ at folds and singular points.
Indeed, if we look at a map $X(\xi)+\delta X(\xi)$ at a point $\xi_0$ on
a fold, then to first order in $\delta X$ the point
$X(\xi_0)+\delta X(\xi_0)$ is on a fold of the new map. In other words
the folds and singular points of the map $X(\xi)+\delta X(\xi)$ are
the locus of the points $X(\xi)+\delta X(\xi)+O(\delta X^2)$ for $\xi$
at a fold of $X(\xi)$.

Let a fold be parametrized in the world-sheet as
$\xi^\alpha=\xi^\alpha(t)$  where
$t$ is a real parameter say in $[0,2\pi)$. In the target space, the
fold is parametrized as $X^\mu = X^\mu(\xi(t))$.
In order to have a well defined measure for maps with folds we have
to correct the metric (\ref{METRDX}) and add terms on folds.
In order to keep the world-sheet diffeomorphism invariance we have
to keep the invariance of the added term under $t\rightarrow \tau(t)$.
The term must thus be of the form
$$
\int G_{\sigma\tau}\delta X^\sigma
       \delta X^\tau\sqrt{\Guv{{d\Xu(\xi(t))}\over{dt}}
    {{d\Xv(\xi(t))}\over{dt}}} \, dt
$$
The integral is thus over the proper length of the fold.
For isolated singular points (including singular points on a fold)
we have to add
$$
\sum_j\Guv\delta\Xu(\xi_j)\delta\Xv(\xi_j)
$$
where $\xi_j$ is the location of the $j$-th singular point.
The corrected measure is thus
\begin{eqnarray}
\| \delta X \|^2 &=& \int \Guv\delta \Xu\delta\Xv\sqrt{\det(\gab)}d^2\xi
\nonumber\\
&+&a\int G_{\sigma\tau}\delta X^\sigma\delta X^\tau
     \sqrt{\Guv{{d\Xu(\xi(t))}\over{dt}}
    {{d\Xv(\xi(t))}\over{dt}}} \, dt \nonumber\\
&+&b \sum_j\Guv\delta\Xu(\xi_j)\delta\Xv(\xi_j)
\label{METRDXC}
\end{eqnarray}

where $a$ and $b$ are arbitrary constants.


In principle we could choose instead of the $b$-term a more complicated
term with a factor that depends on the distances between the singular
points and the lengths of the arcs of a fold between consecutive singular
points on it. However, (\ref{METRDXC}) is the only choice that is
{\em local} in $X(\xi)$.
The $a$-term is actually  the measure for a
relativistic point particle in 2D.

To summarize,
we have argued that the measure in target space for this topological sector
of maps with a single connected fold (which we assume to be homotopic
to a circle) and $k$ singular points
can be taken to be proportional to:
\begin{equation}
{{\prod_t (dX^1(\xi(t))dX^2(\xi(t)))}\over{{\rm Diff}_1(t\mapsto\tau(t))}}
\prod_{i=1}^k ({{dl}\over{dt_i}}dt_i)
\label{MSRFLD}
\end{equation}
where
\begin{equation}
{{dl}\over{dt}}=\sqrt{\Guv{{d\Xu(\xi(t))}\over {dt}}
              {{d\Xv(\xi(t))}\over {dt}}}
\end{equation}
and ${\rm Diff}_1(t\rightarrow\tau(t))$ means that we have to gauge fix this
one-dimensional group of diffeomorphisms along the fold.
The first term in (\ref{MSRFLD}) is exactly the ``Nambu-Goto'' measure for a
bosonic point-particle.

The geometrical meaning of (\ref{MSRFLD}) is as follows.
We know that the full NG measure is
\begin{equation}
{{\prod_\xi (dX^1(\xi)dX^2(\xi))}\over {{\rm Diff}_2 (\xi\mapsto\eta(\xi)).}}
\end{equation}
Because of the diffeomorphism invariance any infinitesimal
change $d\Xu(\xi)$ for a $\xi$ that is not {\em on} the fold (i.e. not
on $\xi(t)$) can be gauged away by a world-sheet coordinate transformation
from ${\rm Diff}_2$. The only changes $d\Xu(\xi)$ that cannot be gauged away
are those that deform the fold in target-space, i.e. those for which
$\xi=\xi(t)$ for some $t$. For these values of $\xi$ the world sheet
diffeomorphism group is reduced to one parameter -- that is, only a change
$d\Xu$ that is parallel to the tangent to the fold at that point, can be
eliminated by a gauge transformation. At the special points $t_i$
any infinitesimal translation of the singular point gives a different target
space picture, and thus cannot be absorbed by a world-sheet diffeomorphism.
Let us decompose
\begin{equation}
dX^1(\xi(t_i))dX^2(\xi(t_i)) = dl_i dy_i
\end{equation}
where $dl_i$ is a translation of the singular point in a direction parallel
to the fold, and $dy_i$ is a translation perpendicular to the fold (that
cannot be removed by a world-sheet diffeomorphism). The part $dy_i$
contributes to the measure of the fold (modulo ${\rm Diff}_1$) while $dl_i$
is the measure for the location of the singular point on the fold.
We thus obtain (\ref{MSRFLD}).

The theory can be rigorously defined on a discrete target space.
We will pick a triangulation of the target space and define the
discrete configuration space of the theory as all the possible ways
to associate two integer cover-numbers with each elementary cell
of the triangulation, and to specify how the sheets join each other
on the edges where two cells touch.
In this way we bypass the need for a world-sheet and do not have
any reparameterization invariance.
This is different from the continuum version in the fact that
singular points of all types (branch points, singular points on folds,
or spiral points) can exist only on the vertices of the triangulation
and folds can exist only on the edges.
The discrete maps are combinatorial objects.
If we restrict the configuration space of all such combinatorial
maps, to those that have no spiral points and no more than one fold
on every edge, we will get a lattice version of the measure (\ref{MSRFLD}).

\subsection{Invariance under area preserving diffeomorphisms}
As was mentioned in the previous section, the NG action is invariant
under APD of the target-space.
The measure $\Dcal X$ seems also invariant since $\det(\gab)$
in (\ref{METRDX}) is APD invariant and the determinant of the metric
in the space of maps is thus APD invariant.
However,  it is easy  to realize
 that because of the $a$-term and the ``proper'' length along the
fold, (\ref{METRDXC}) is {\em not APD invariant}.

The term $\prod_t( dX^1(t) dX^2(t))$ which appears
in the measure as a consequence of the $a$-term in (\ref{METRDXC})
is invariant under (target-space)
APD, however, the measure has the additional factor of ${{ds}\over{dt}}$
where $ds$ is the proper time, and this breaks APD invariance.

It is interesting to note
 that there exists an area preserving
diffeomorphism of the target space that changes the fold $Y^\mu(t)$,
where $t\in [0,2\pi)$ is parameterizing the fold, to
any other form $Z^\mu(t)$ provided it has the same area.

We will show it for infinitesimal changes $\delta Y^\mu(t)$ that satisfy
\begin{equation}
\int \epsilon_{\mu\nu}\delta Y^\mu(t) \dot{Y}^\nu(t) dt=0
\label{LOOPC}
\end{equation}
representing the fact that the area is unchanged.
We want to extend $\delta Y^\mu(t)$ to a function on the whole of target
space $u^\mu(X)$ such that
\begin{eqnarray}
u^\mu(Y(t))&=&\delta Y^\mu(t) \\
\partial_\mu u^\mu(X) &=& 0
\end{eqnarray}
The second identity represents the invariance of $\sqrt{g}$.
We can seek a solution in the form
\begin{equation}
u^\mu(X)=\epsilon^{\mu\nu}\partial_\nu\phi(X)
\end{equation}
which is consistent with (\ref{LOOPC}). Clearly $\phi(X)$ can be chosen
so as to satisfy the boundary condition on the folds.

\subsection{World-Sheet Reparametrization Gauge Fixing}

Gauge fixing of the world-sheet diffeomorphisms requires a special care
in string theory with folds. This is demonstrated in the use of
the conformal gauge. The latter means inserting the delta functions
$\prod_\xi\delta(g_{++}(\xi))\delta(g_{--}(\xi))$ together with the
corresponding Fadeev-Poppov determinant. However, this procedure is not
applicable when the map $X$ has folds. The reason is that on folds the
map $X$ is singular, and thus $\det(\gab)=0$. In the conformal gauge
this implies that $\gab=0$ on folds, and thus, since the perimeter
of a fold is greater than zero, the fold must lie at the infinity of
the $\xi$ parameter space. Using the world-sheet light-cone gauge is also
not adequate since inherently in that gauge $\det(\gab)=1$.
One can try to overcome this problem by, for instance, fixing
$g_{++}(X)=g_{++}^0$ where $g_{++}^0$ is some nonzero constant value, (and
similarly $g_{--}(X)=g_{--}^0$), however this will prevent the map $X$ from
having branch points at which $\gab=0$.

\section{The contribution of folded maps to the partition function}
In the following sections we will concentrate on the consequences
of the $b$-term in (\ref{METRDXC}). Our approach will be to sort the
maps into sectors. Each sector will be specified by the way in which
the folds divide the target-space. More precisely, it will be specified
by the graph that is formed from the smooth regions (as vertices of the
imaginary graph) and the folds (as edges connecting the smooth regions).

We then have to sum over all possible number of branch points inside
the smooth regions, and all possibilities of singular points on the
folds. The measure for the latter will be given by the
$b$-term of (\ref{METRDXC}). The summations are essentially of a
{\em combinatorial} nature, because we have to consider various
ways in which the sheets of the smooth regions join.

After we have done this, we should,  in principle, integrate over all
the maps of the sector. This involves an integration over the
{\em location} of the folds and is given by the $a$-term of
 (\ref{METRDXC}).
In contrast to the $b$-term contribution, the $a$-term contribution
is of a {\em dynamical} nature. It involves integration over
self-avoiding random walks.

In section (2.2) we studied the {\em classical} possibilities for singular
points in a map. Our assumption was that the map function $\Xu(\xia)$ can be
Taylor expanded and that the maps are generic.
In particular we saw that the form of singular points on a fold is very
restricted and, for example, a singular point on a fold, for a {\em generic}
map cannot have a branch cut starting from it. ``Spiral-points'' (as in
Fig.7) are excluded.

\putfig{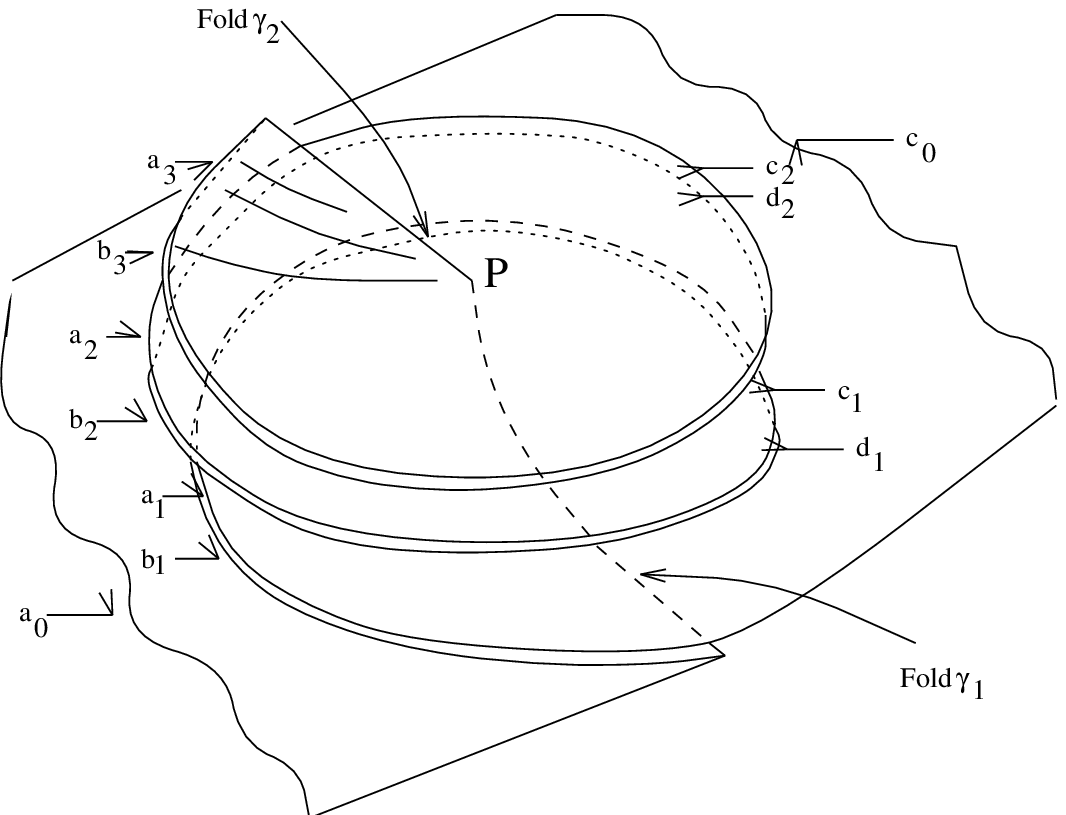}{Fig. 7:
 \mbox{$P$} is a spiral point of order \mbox{$2{1\over 2}$}.
 along \mbox{$\gamma_1$}:
  \mbox{$a_0$} and \mbox{$b_1$} are joined by the fold and the
 smooth passages are: \mbox{$a_1-c_0$},
 \mbox{$a_2-c_1$}, \mbox{$a_3-c_2$}, \mbox{$b_2-d_1$}, \mbox{$b_3-d_2$}.
 Along \mbox{$\gamma_2$}: \mbox{$a_3$} and \mbox{$b_3$}
 are joined by the fold and the
 smooth passages are: \mbox{$a_0-c_0$},
 \mbox{$a_1-c_1$}, \mbox{$a_2-c_2$},
  \mbox{$b_1-d_1$}, \mbox{$b_2-d_2$}.}{100mm}

\putfig{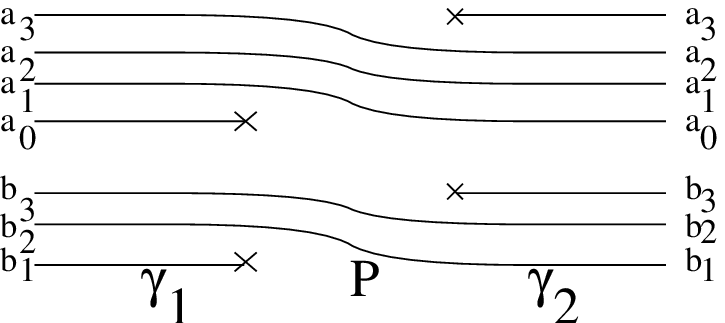}{Fig. 8: A graphic representation
    of the connection of the sheets near \mbox{$P$}.}{60mm}

\putfig{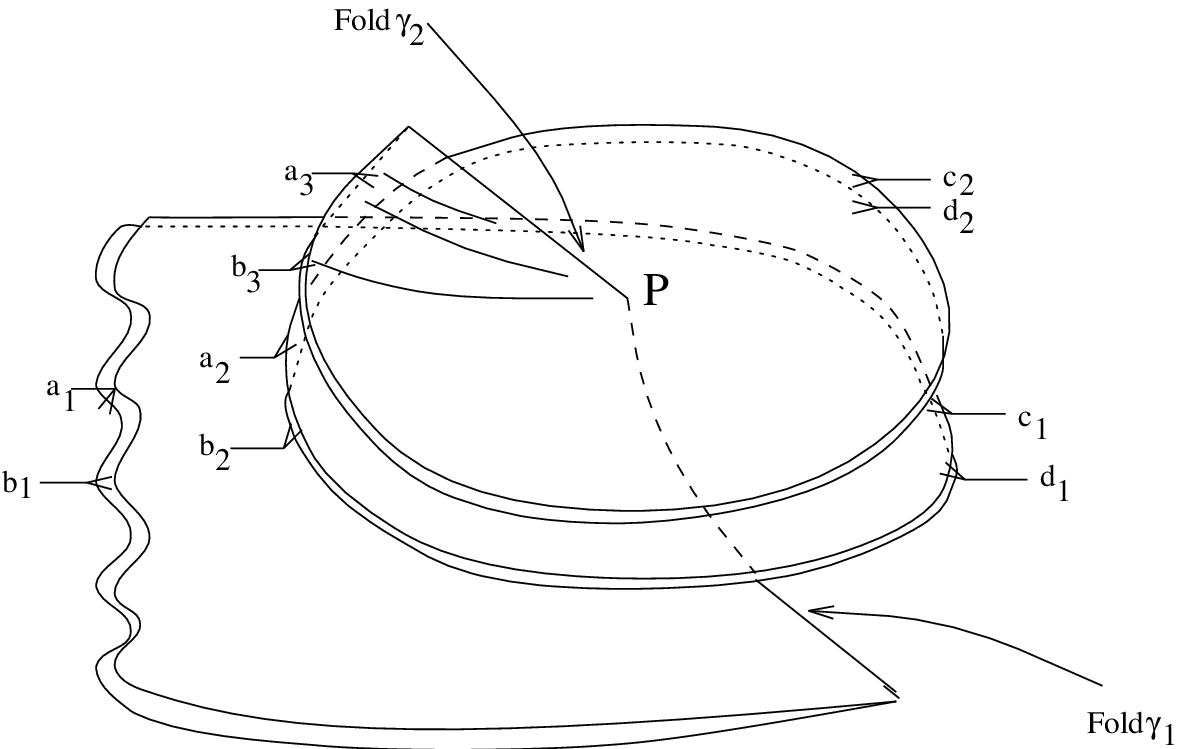}{Fig. 9:
 \mbox{$P$} is a spiral point of order \mbox{$2$}.
 Along \mbox{$\gamma_1$}:
  \mbox{$a_1$} and \mbox{$b_1$} are joined by the fold and the
 smooth passages are: \mbox{$a_2-c_1$}, \mbox{$a_3-c_2$},
 \mbox{$b_2-d_1$}, \mbox{$b_3-d_2$}.
 Along \mbox{$\gamma_2$}:
 \mbox{$a_3$} and \mbox{$b_3$} are joined by the fold and the
 smooth passages are: \mbox{$a_1-c_1$}, \mbox{$a_2-c_2$},
 \mbox{$b_1-d_1$}, \mbox{$b_2-d_2$}.}{100mm}

\putfig{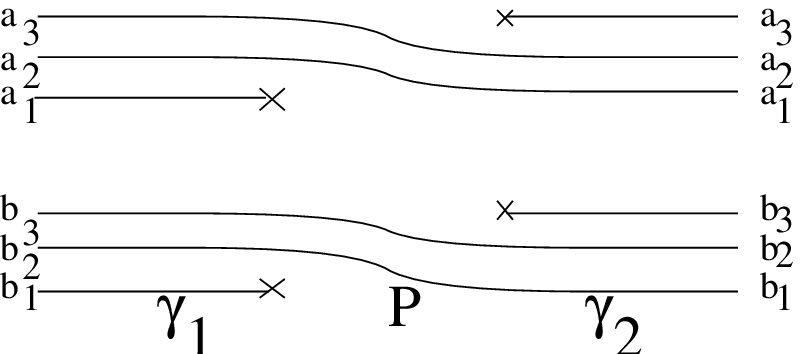}{Fig. 10:
A graphic representation
 of the connection of the sheets near \mbox{$P$}.}{60mm}

These arguments probably do not persist at the quantum level.
Renormalizability arguments probably force us to include additional special
points like ``spiral-points''.
 When we take a cutoff $\Lambda$, of dimension of
mass, to regularize the theory we allow a branch point to be close to  a
fold only by a distance of the order of $\Lambda^{-1}$. All configurations
with branch points that are closer than $\Lambda^{-1}$ to a fold should be
considered as configurations with branch points {\em on} the fold.
classically, the ``volume'' of such configurations (in map space) tends to
zero. However, renormalization might give it a nonzero weight.
Similarly, infinitesimal contracted handles and connecting tubes which
were needed in \cite{GT2} might arise by a similar mechanism.

We will begin a study of the simple cases that arise, not restricting
ourselves to the ``classical'' configurations.
To examine the role played by the folds we start with a simple example
of maps from a sphere to a sphere with winding number one and a single fold.
Suppose that an area enclosed by the arcs $\gamma_1$ and $\gamma_2$ in
$\Sigma$ is mapped by $X$ to a closed region $A$ in $\tsp$ in such a
way that on $\tsp$ the region $A$ is a 3-sheet cover and $\tsp-A$ is a
single-sheet cover. The region $A$ thus forms a pocket (see Fig.2).
The two curves $\gamma_1,\gamma_2$ that mark the folds are mapped into
$\gamma_1',\gamma_2'$. If we pick a random labeling of the
sheets of $A$ calling them $a,b,c$ then at the boundary $\gamma_1'$
the sheets $a,b$ are glued together and at $\gamma_2'$ the sheets
$c,b$ are joined together, thus forming a ``pocket".

The next topological situation (see Fig.11)
we would like to examine is a region $A$
in $\tsp$ that is a 3-sheet cover bounded by $k$ arcs
$\gamma_1,\gamma_2,\dots,\gamma_k$ (we drop the prime from arcs on the
target space). We will assume that the outside region $\tsp-A$ is a single
sheet.
Along each $\gamma_i$ two of the three sheets join.
We require that at $\gamma_i$ and $\gamma_{i+1}$ the joining of
the sheets is different ($k+1$ is by definition $1$).
There are altogether 3 different ways to join two sheets out of three.
However two of the sheets have orientability $(+)$, like the outer single
sheet and the last sheet has orientability $(-)$.
So, if we want an oriented figure, we are allowed
to join only a $(+)$ sheet to a $(-)$ sheet. This leaves only two ways to
join, and since adjacent $\gamma_i$-s must not have the same joining,
there are altogether two ways to join the sheets. Since the labeling
of the two $(+)$ sheets was arbitrary, there is a symmetry factor of 2,
which cancels out -- leaving only one way to join.

\putfig{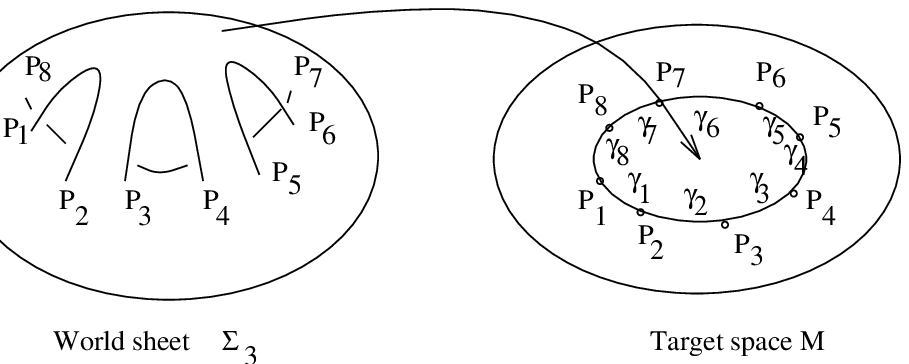}{Fig. 11: A map from a world sheet of high genus to
                            a sphere.}{100mm}

The Euler number of $\Sigma$, $\chi(\Sigma)$, is given by
$$\chi(\Sigma)={\rm Vertices} + {\rm Faces} - {\rm Edges} = 4-k$$
When we consider the partition function of the string we will have to sum
over all positions of the $k$ endpoints of the $\gamma_i$.
Such configurations should thus get a factor of $(\rho l)^k/k!$
where $l$ is the length of the boundary of $A$
and $\rho$ is a parameter of dimension $[Length]^{-1}$ that is adjoined
with singular points.

Clearly $k$ must be even and
in order that the world-sheet should be  connected we need $k\geq 2$.

Summing over even $k$ one finds

\begin{eqnarray}
e^{Z(\tsp)}&=&e^{-\mu \cdot Area(\tsp) - 2\mu\cdot Area(A)}
    \sum_{g=0}^\infty \lmst^{2g-2} {{(\rho l)^{2g+2}}\over{(2g+2)!}}\nonumber\\
 &=& {1\over {2\lmst^4}} e^{-\mu \cdot Area(\tsp) - 2\mu\cdot Area(A)}
      (e^{\lmst\rho l} + e^{-\lmst\rho l})
\label{ZSING0}
\end{eqnarray}

Notice that we included the term $k=0$ which corresponds to a disconnected
world-sheet. We thus calculate $e^{Z(\tsp)}$ rather than $Z(\tsp)$.

Equation (\ref{ZSING0}) suggests a relation between a fold and a
trajectory of a point particle.  The action of 	a point particle
coupled to a YM $SU(N)$ field is
proportional to the length of its trajectory times its mass plus
the expectation value of the Wilson loop around its trajectory
which is proportional to the area if the trajectory is not
self intersecting
\cite{STROMING}. According to this interpretation, (\ref{ZSING0})
is a sum of two terms which correspond to a particle of mass
$\lmst\rho$ and a ghost particle of negative mass $-\lmst\rho$.
After we perform a Wick rotation we get an imaginary $\rho$ and a real mass
$\pm\lmst |\rho|$. The negative mass does not create a problem if only
the square of the mass appears in final results.
However, the interpretation of a massive particle, o
is not consistent when we go to
higher cover numbers, as we shall discuss later.

\subsection{The Fold Transition Matrix}
In the previous section we considered folded maps based on a simply connected
 region
covered with two sheets of positive orientability $(+)$ and one of negative
orientability $(-)$. The region was surrounded,
in the rest of the target space, by a single cover $(+)$ sheet.
We generalize this simple set of maps to maps
with a simply connected  region that has $n+1$ sheets of orientability $(+)$
and $m+1$ sheets of orientability $(-)$, whereas the outside region has
$n$ sheets of orientability $(+)$ and $m$ sheets of orientability $(-)$.

Again, we pick $k$ singular points along the curve $\gamma$,
$\gamma = \bigcup_{i=1}^k\gamma_k$ that bounds  the region $A$.
Along each $\gamma_i$ we perform the following operations:
\begin{enumerate}
\item
  join one of the $n+1$ orientability $(+)$ sheets to one of the
  $m+1$ orientability $(-)$ sheets,
\item
  connect the $m$ remaining orientability $(-)$ sheets of the inside to
  the $m$ remaining orientability $(-)$ sheets of the outside in some random
  permutation,
\item
  connect the $n$ remaining orientability $(+)$ sheets of the inside to
  the $n$ remaining orientability $(+)$ sheets of the outside in some random
  permutation.
\end{enumerate}

The genus of the obtained figure can be determined as before from Euler's
formula. Assuming that the $n+m$ sheets of
the outside region are disconnected (in $\tsp-A$) we get:
\begin{eqnarray}
{\rm Vertices} &=& (m+n)k +\sum({\mbox spiralities})
\nonumber \\
{\rm Faces} &=& 2n+2m+2 \\
{\rm Edges} &=& (n+m+1)k \nonumber
\end{eqnarray}
The term
\footnote{We thank W. Taylor for correcting us with regards to this formula.}
$\sum({\mbox spiralities})$ is related to points of the type of
Fig.7  or Fig.9 .
In the ``classical'' configurations, as we saw earlier,
no such points exist and we obtain that
the sum of the Euler numbers is:
\begin{equation}
\sum_{components} (2-2g_i) = 2n+2m+2-k   \label{GENUSNMK}
\end{equation}
We see that each singular point contributes $(-1)$
to the Euler characteristic,
and that every cover contributes $(+1)$.

Let $(x_i,y_i)$ for $i=1,\dots,k$  be the sequential numbers of the joined
sheets at $\gamma_i$. Thus $1\leq x_i\leq m+1$ and $1\leq y_i \leq n+1$.
Let $\sigma_i\in S_{n+1}$ be defined as follows:
$\sigma_i(x)$ for $x=1,2,\dots,n$ will be the sheet number from
$1,2,\dots,n+1$ to which $x$ continues as it crosses the fold,
and $\sigma_i(n+1)$ is defined to be $x_i$.
Likewise we define $\tau\in S_{m+1}$ as follows:
$\tau_i(y)$ for $y=1,2,\dots,m$ will be the sheet number from
$1,2,\dots,m+1$ to which $y$ continues as it crosses the fold,
and $\tau_i(m+1)$ is defined to be $y_i$.

The sequence of permutations $\{\sigma_i\}_{i=1}^k,\{\tau_i\}_{i=1}^k$
characterizes the topology of the figure precisely.
This is up to a symmetry factor which is related to the arbitrariness in
the labeling of the $n+1$ sheets and the $m+1$ sheets, as well as the
possibility to relabel the outer $n$ and $m$ sheets.
To be more precise, it is (up to a factor) the number of permutations
$\psi\in S_{n+1}$ and $\phi\in S_{n}$ such that
$\sigma_i=\psi\sigma_i\phi$ for all $i$, times the number of permutations
$\theta\in S_{m+1}$ and $\beta\in S_m$ such that
$\tau_i=\theta\tau_i\beta$ for all $i$.

In general , when we pass from $\gamma_i$ to $\gamma_{i+1}$ not all the
configurations $(\sigma_i,\tau_i)\rightarrow (\sigma_{i+1},\tau_{i+1})$
will be allowed. For example,
we should require that $(x_{i+1},y_{i+1})\neq(x_i,y_i)$
as an ordered pair.
If $(x_{i+1},y_{i+1})=(x_i,y_i)$ there is still the possibility to make
 a nontrivial permutation in the gluing of the inner sheets to the outer
 sheets, however this topological property cannot be assigned to a
 fold, i.e. a singularity along a line.

Furthermore, for general string theories (defined by the coupling constants)
there will be different weights for different
$(\sigma_i,\tau_i)\rightarrow (\sigma_{i+1},\tau_{i+1})$.

Let us denote by $F^{\sigma_1\tau_1}_{\sigma_2\tau_2}$ the weight that
is attributed to the passage
$(\sigma_i,\tau_i)\rightarrow (\sigma_{i+1},\tau_{i+1})$
along the fold. $F$ is a $(n+1)!(m+1)!\times(n+1)!(m+1)!$ symmetric
matrix, which we shall call {\em the fold-matrix}.

The factor for a fold with $k$ singular points (and with the assumption that
inside the area $A$ the boundaries of the $(n+1)+(m+1)$ sheets are
disconnected is:
\begin{equation}
\sum_{\sigma_i,\tau_i} S(\{\sigma_i\}_{i=1}^k,\{\tau_i\}_{i=1}^k)
 F^{\sigma_1\tau_1}_{\sigma_2\tau_2}F^{\sigma_2\tau_2}_{\sigma_3\tau_3}
 \dots F^{\sigma_k\tau_k}_{\sigma_1\tau_1}
\label{FOLDEQ}
\end{equation}
where the symmetry factors are 1 over the number of equivalence classes of
$\{\sigma_i\}_{i=1}^k$, $\{\tau_i\}_{i=1}^k$. It can be recast as
\begin{eqnarray}
S(\{\sigma_i\}_{i=1}^k,\{\tau_i\}_{i=1}^k) &=&
{1\over{(n+1)!(m+1)!n!m!}} \nonumber\\
&\cdot& \sum_{\psi\in S_{n+1},\phi\in S_n}
(\prod_{i=1}^k \delta(\sigma_i^{-1}\psi\sigma_i\phi)) \nonumber\\
&\cdot& \sum_{\theta\in S_{m+1},\beta\in S_m}
(\prod_{i=1}^k \delta(\tau_i^{-1}\theta\tau_i\beta))
\label{SYMMFAC}
\end{eqnarray}
where $\delta$ is 1 if and only if its argument is the identity permutation.
Thus the factor for $k$ singular points is:
\begin{equation}
{1\over{(n+1)!(m+1)!n!m!}}\sum_{\psi,\phi,\theta,\beta}
Tr (F_{(\psi,\phi,\theta,\beta)}(n,m)^k)
\end{equation}
where:
\begin{eqnarray}
F_{(\psi,\phi,\theta,\beta)}(n,m)
&\stackrel{def}{=}& F^{\sigma_1\tau_1}_{\sigma_2\tau_2}
  \delta(\sigma_1^{-1}\psi\sigma_1\phi))
  \delta(\sigma_2^{-1}\psi\sigma_2\phi)) \nonumber\\
 &\cdot&  \delta(\tau_1^{-1}\theta\tau_1\beta))
         \delta(\tau_2^{-1}\theta\tau_2\beta))
\label{RESTRF}
\end{eqnarray}
We will call the $F_{(\psi,\phi,\theta,\beta)}(n,m)$-s the
{\em restricted fold matrices}.

The formula for the weight of a fold of length $l$ becomes:
\begin{eqnarray}
W &=& \sum_{k=0}^\infty \lmst^{k-2(n+m+1)} {{l^k}\over{k!}}
\sum_j\sum_{\begin{array}{c}\psi\in S_{n+1},\phi\in S_n \\
                            \theta\in S_{m+1},\beta\in S_m
            \end{array}}
(\alpha_{(\psi,\phi,\theta,\beta),j}(n,m))^k
    \nonumber\\
&=&  \sum_j\!\!
      \sum_{\begin{array}{c}\psi\in S_{n+1},\phi\in S_n \\
                            \theta\in S_{m+1},\beta\in S_m
            \end{array}}\!\!
\lmst^{-2(n+m+1)}
    e^{\lmst\alpha_{(\psi,\phi,\theta,\beta),j}(n,m)\, l} \nonumber\\
&&
\label{FOLDSP}
\end{eqnarray}
where  $\alpha_{(\psi,\phi,\theta,\beta),j}(n,m)$ are the eigenvalues of the
restricted fold-matrices.
We have used (\ref{GENUSNMK}) and absorbed the ``spiralities''  into the
definition of the fold-matrix.
The determination of the weight of the fold thus reduces to the determination
of the eigenvalues of the
restricted fold-matrices.
In appendix A we will calculate the fold matrix for general
$(n,m)$ under the assumption of no branch points on the folds (no spiralities).
In appendix B we will show how to calculate the
eigenvalues of the restricted fold matrices of appendix A.

We will end this subsection with a description of
the slight modification that
is needed when the boundaries of the $(n\!+\!1)+(m\!+\!1)$
sheets are not disconnected,
as is the case when there are branch-points inside $A$ that do not cancel
each other.  When one goes
along one sheet close to the fold boundary he will not return to the same
sheet after one whole round, because there are branch cuts that start at
the branch points inside the area bounded by the fold.
Since the branch cuts can be defined arbitrarily, so long as they start at
the branch points, we can make all the branch cuts pass through the fold
segment between the last singular point and the first one.
Note that when branch cuts are present singular points must exist
when the fold is non-intersecting. Otherwise the fold
will not close well.
We will define permutations $\zeta\in S_{n+1}$ and
$\xi\in S_{m+1}$ that describe how the $(n\!+\!1)$
(respectively $(m\!+\!1)$)
sheets change to one another as one finishes a full round close to the
boundary of $A$. In \cite{GT2} the equivalence classes of $\zeta$ and
$\xi$ were related to the representation in which a Wilson loop operator
was taken around the boundary.

Eqn.(\ref{FOLDEQ}) is  now modified to
\begin{equation}
\sum_{\sigma_i,\tau_i} S(\{\sigma_i\}_{i=1}^k,\{\tau_i\}_{i=1}^k)
 F^{\sigma_1\tau_1}_{\sigma_2\tau_2}F^{\sigma_2\tau_2}_{\sigma_3\tau_3}
 \dots F^{\sigma_k\tau_k}_{(\zeta\sigma_1)(\xi\tau_1)}
\label{FOLDEQ1}
\end{equation}
This can still be recast as a sum of eigenvalues if we write the product as
$tr(F^k P(\zeta,\xi))$ where the matrix $P$ is given by
\begin{equation}
P(\zeta,\xi)^{\sigma_1\tau_1}_{\sigma_2\tau_2}
= \delta(\zeta\sigma_1\sigma_2^{-1})\delta(\xi\tau_1\tau_2^{-1})
\end{equation}
After diagonalizing the restricted fold matrices, the factor of the
fold can still be written as a sum over exponents like (\ref{FOLDSP}) but
with a factor in front of each exponent which is the diagonal element of
$P$ in the basis in which the restricted fold matrix is diagonal.

We have to include also the effect of branch points in $\tsp-A$.
(i.e. a nontrivial boundary for $\tsp-A$ near but a little bit to the
outside of A).
This can be done similarly, by introducing permutations.
We will define permutations $\zeta'\in S_n$ and
$\xi'\in S_m$ that describe how the $n$ (respectively $m$)
sheets change to one another as one finishes a full round close to the
boundary of $\tsp-A$. We then have to insert
\begin{equation}
P'(\zeta',\xi')^{\sigma_1\tau_1}_{\sigma_2\tau_2}
= \delta(\sigma_1\zeta'\sigma_2^{-1})\delta(\tau_1\xi'\tau_2^{-1})
\label{PZXDEF}
\end{equation}

\subsection{Quartic Vertices}
The next step after considering the contribution to the partition function
of a single fold is to consider the situation of several (possibly
intersecting) folds.
The figurative picture of a general map with folds is  that of regions in the
target space that are bounded by folds (see Fig.1 for example).
We wish to develop ``Feynman rules'' to obtain the weight of such
diagrams. Generic maps have folds, that are curves which intersect
at certain points in which one fold passes above the other fold.
Such points will be quartic vertices in the Feynman rules.
Another source of quartic vertices is the quartic points of (\ref{QUARTP}).

In this subsection we  obtain the contribution of the vertex and in
the next subsection we will write down the full set of Feynman rules
(in a formal form).

\putfig{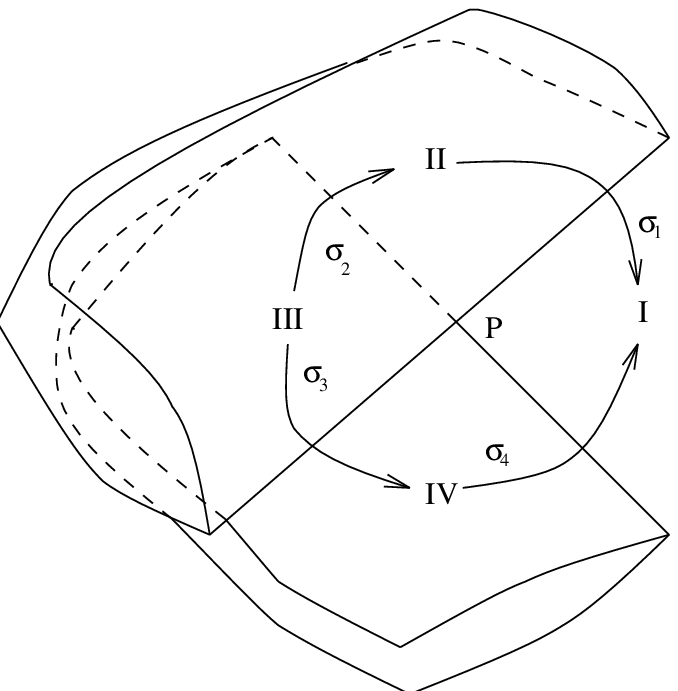}{Fig. 12: A point where one fold passes on top of
                       the other.}{80mm}

Let $P$ be a point at which one fold passes on top of the other (Fig.12).
We will assume that the point $P$ itself is not a singular point on a fold,
and thus we can restrict ourselves to a small neighborhood of $P$ in which
there are no singular points on the folds. Near $P$ the folds divide the
space into four regions with cover numbers $(n,m)$,$(n+1,m+1)$,
$(n\!+\!1,m\!+\!1)$, $(n\!+\!2,m\!+\!2)$.
Let $\sigma_1\in S_{n+1}$ be the permutation that specifies
how the $n$ sheets from region $I$ are continued to the sheets of region
$II$ (for some pre-determined labeling). We recall our convention that
$\sigma_1(n+1)$ is the sheet in region $II$ that is connected to the fold.
similarly we define $\sigma_4\in S_{n+1}$ for the passage $I\rightarrow IV$
and $\sigma_3,\sigma_2\in S_{n+2}$ for the passages $IV\rightarrow III$ and
$II\rightarrow III$ respectively (see Fig.12).
 We define the permutations
$\tau_1,\tau_2,\tau_3,\tau_4$ for the covers with the opposite orientation,
in a similar fashion. Our requirement that $P$ be an ordinary point
translates into the requirement that
\begin{equation}
\sigma_4^{-1}\sigma_3^{-1}\sigma_2\sigma_1 = P_{n+1,n+2}
\end{equation}
where $P_{n+1,n+2}\in S_{n+2}$ is the permutation that switches
$(n\!+\!1)$ and $(n\!+\!2)$.
The reason for this relation is as follows.
After we make one loop around $P$,
covers $1,\dots,n$ in region $I$ return to themselves
(otherwise $P$ would be a branch point).
When we start with (the formal) cover number $(n\!+\!1)$
 in region $I$ we surround the point
$P$ in the following manner (see Fig.12):
first we go over to the cover that is attached
 to the fold in region $IV$, we pass region $III$,
 we cross over to region $II$ and then we meet the fold.
So, by our conventions we end up with (the formal) cover number $(n\!+\!2)$.
Similarly, for the $\tau$-s we have
\begin{equation}
\tau_4^{-1}\tau_3^{-1}\tau_2\tau_1 = P_{m+1,m+2}
\end{equation}

The quartic points of (\ref{QUARTP}), on the other hand, satisfy either
\begin{eqnarray}
\sigma_4^{-1}\sigma_3^{-1}\sigma_2\sigma_1 &=& P_{n+1,n+2} \nonumber\\
\tau_4^{-1}\tau_3^{-1}\tau_2\tau_1 &=& (1)
\end{eqnarray}
or
\begin{eqnarray}
\sigma_4^{-1}\sigma_3^{-1}\sigma_2\sigma_1 &=& (1) \nonumber\\
\tau_4^{-1}\tau_3^{-1}\tau_2\tau_1 &=& P_{m+1,m+2}
\end{eqnarray}
where $(1)$ is the identity permutation.
These relations can be checked from (Fig.6)
which corresponds to (\ref{QUARTP}).

In the following section we describe a set of rules to calculate the
weight that goes with a {\em given} division of the target-space into
regions. The location of the folds is thus assumed to be fixed.
In these rules we assume that no more than two folds intersect at
any point of the target-space.
In other words, if folds are considered as ``propagators'' of Feynman
diagrams, then there are only quartic vertices and no higher ones.
This is in agreement with the classical discussion in section (2.1), where
it was claimed that a generic map has at most quartic singularities.
We remark, however, that quantum-mechanically, ``contact-terms'' may arise
and thus higher vertices, with more than two folds, may have to be
considered.

\subsection{Feynman Rules}
The full weight of a configuration of folds (as in Fig.1 for example) is
given by the following ``Feynman Rules'':
\begin{enumerate}
\item {\em cover numbers -- }
associate with each region non-negative integer cover-numbers $(n,m)$ in all
possible ways, and such that when crossing each fold both $m$ and $n$ either
increase by $1$ or decrease by $1$.

\item {\em vertices -- }
near each vertex $P$ with folds $\gamma_1$,$\gamma_2$,$\gamma_3$,$\gamma_4$
such that $\gamma_1$ and $\gamma_2$ bound a region $A$ with cover-numbers
$(n,m)$, $\gamma_2$ and $\gamma_3$ bound a region $B$ with cover-numbers
$(n+1,m+1)$, $\gamma_3$ and $\gamma_4$ bound a region $C$ with cover-numbers
$(n+2,m+2)$, and $\gamma_4$ and $\gamma_1$ bound a region $D$ with
cover-numbers $(n+1,m+1)$
($B$ and $D$ are not necessarily different regions),
introduce the following variables:
 $\sigma_1\in S_{n+1}$ and $\tau_1\in S_{m+1}$ on the fold
$\gamma_1$, $\sigma_2\in S_{n+1}$ and $\tau_2\in S_{m+1}$ on
the fold $\gamma_2$, $\sigma_3\in S_{n+2}$ and $\tau_3\in S_{m+2}$ on
the fold $\gamma_3$, and $\sigma_4\in S_{n+2}$ and $\tau_4\in S_{m+2}$
on the fold $\gamma_4$. Adjoin a factor of
\begin{eqnarray*}
\lefteqn{
\delta(\sigma_4^{-1}\sigma_3^{-1}\sigma_2\sigma_1 P_{n+1,n+2})
\delta(\tau_4^{-1}\tau_3^{-1}\tau_2\tau_1 P_{m+1,m+2})
} \\
&+&\lambda_{qp}
\delta(\sigma_4^{-1}\sigma_3^{-1}\sigma_2\sigma_1)
\delta(\tau_4^{-1}\tau_3^{-1}\tau_2\tau_1 P_{m+1,m+2}) \\
&+&\lambda_{qp}
\delta(\sigma_4^{-1}\sigma_3^{-1}\sigma_2\sigma_1 P_{n+1,n+2})
\delta(\tau_4^{-1}\tau_3^{-1}\tau_2\tau_1)
\end{eqnarray*}
where $\lambda_{qp}$ is the weight of quartic points as in
(\ref{QUARTP}) and is a free parameter of the model.
The first term is the contribution of configurations where
one fold just passes (without touching) on top of the other fold.
We stress that renormalization may change the form of quartic vertices,
and may add 6-vertices, 8-vertices and so on.

\item {\em smooth regions -- }
each smooth region gets a factor of $e^{-\mu A}$ where $A$ is the area
of the region and $\mu$ is the cosmological constant.
In addition, we can include branch points, collapsed handles and
contracted tubes each with factors of $\lambda_{br}A$, $\lambda_{ch}A$
and $\lambda_{ct}A$ where the $\lambda$-s are the corresponding coupling
constants. Also every insertion
 contributes a corresponding power of $\lambda_{st}$ - the string
coupling constant, according to the change in world-sheet genus.
An additional complexity that is caused by the branch points is the fact
that the boundary of the region that contains branch points does no longer
have to be made of $n$ (and $m$) distinct loops, but as we make a full
loop around the region we may end up on a different sheet
from the one we started with.
Let $\xi\in S_n$ and $\zeta\in S_m$ denote the permutations that
represent the changing of the sheets as we make a full round.
These $\xi$-s and $\zeta$-s will have an effect on the symmetry-factor
considerations and on the propagators.

\item {\em symmetry factors -- }
adjoin with each smooth region of covers $(n,m)$ two permutations
$\psi\in S_n$ and $\theta\in S_m$ that commute with the $\xi$ and $\zeta$
that where associated with that region if it has branch points (otherwise
the $\xi$ and $\zeta$ can be taken as the identity).

\item {\em propagators -- }
give each fold a factor of
\begin{equation}
(\exp(\lmst l
F_{(\psi,\phi,\theta,\beta)}(n,m)))^{\sigma_1\tau_1}_{\sigma_2\tau_2}
\end{equation}
where $F$ is the restricted fold matrix (see (\ref{RESTRF})),
$\psi$ and $\theta$ are the symmetry-factor permutations associated with the
region on one side of the fold, and $\phi$ and $\beta$ are those associated
with the other side of the fold. $(n,m)$ are the cover numbers of the
region with the fewer covers. $\sigma_1$ and  $\tau_1$ are the permutations
that were written on the fold near the first vertex that bounds it and
$\sigma_2$ and $\tau_2$ were written on the fold near the second vertex that
bounds it. $l$ is the length of the fold and $\lmst$ is the string coupling
constant.

\item {\em branch cuts -- }
 There is a further complication because of the $\xi$-s and
$\zeta$-s of the smooth region. For each smooth region we must pick
one propagator (i.e. fold) that bounds it, and insert a change of the
a labeling of the covers of the region at a certain point.
This is done by replacing
$(\exp(\lambda l
F_{(\psi,\phi,\theta,\beta)}(n,m)))^{\sigma_1\tau_1}_{\sigma_2\tau_2}$
for that propagator, with
$$
(\exp(\lambda l_2 F_{(\psi,\phi,\theta,\beta)}(n,m))
P(\xi,\zeta)\exp(\lambda l_1
F_{(\psi,\phi,\theta,\beta)}(n,m)))^{\sigma_1\tau_1}_{\sigma_2\tau_2}
$$
where $l_1$ and $l_2$ are the lengths from the vertices to the
point where the cut intersects the fold.
The matrix $P$ is given by
$$
P(\zeta,\xi)^{\sigma_1\tau_1}_{\sigma_2\tau_2}
= \delta(\zeta\sigma_1\sigma_2^{-1})\delta(\xi\tau_1\tau_2^{-1})
$$
when $P$ corresponds to a permutation due to a cut from branch points
from the side of the fold where there are more sheets, and is given by
$$
P(\zeta',\xi')^{\sigma_1\tau_1}_{\sigma_2\tau_2}
= \delta(\sigma_1\zeta'\sigma_2^{-1})\delta(\tau_1\xi'\tau_2^{-1})
$$
when $P$ corresponds to a permutation due to a cut from branch points
{}From the side of the fold where there are less sheets.

\item {\em statistics of closed loops -- }
our remark after formula (\ref{GENUSNMK}) about the contribution of the
covers to the genus means that we have to add a factor of $\lmst^2$
for each closed loop (of folds).

\item {\em taking the trace -- }
sum over all permutations $\sigma_i$,$\tau_i$ and the
permutations $\psi$-s and $\theta$-s.
\end{enumerate}

\parbox[c]{150mm}{
\begin{picture}(140,150)
\put (90,120){\line(1,0){30}}
\put (80,120){$\sigma_1,\!\tau_1$}
\put (121,120){$\sigma_2,\!\tau_2$}
\put (100,115){$(n,m)$}
\put (95,125){$(n\! +\! 1,m\! +\! 1)$}
\put (100,110){$(\phi,\beta)$}
\put (100,130){$(\psi,\theta)$}
\put (20,120){$(e^{\lambda l F_{(\psi,\phi,\theta,\beta)}(n,m)})^{
         \sigma_1,\tau_1}_{\sigma_2,\tau_2}$}
\put (90,95){\line(1,-1){30}}
\put (80,98){$\sigma_3,\!\tau_3$}
\put (121,60){$\sigma_1,\!\tau_1$}
\put (90,65){\line(1,1){30}}
\put (80,60){$\sigma_2,\!\tau_2$}
\put (121,98){$\sigma_4,\!\tau_4$}
\put (94,95){$(n\! +\! 2,m\! +\! 2)$}
\put (100,65){$(n,m)$}
\put (109,80){$(n\! +\! 1,m\! +\! 1)$}
\put (86,81){$(n\! +\! 1,$}
\put (90,78){$m\! +\! 1)$}
\put (0,85){$\delta(\sigma_4^{-1}\sigma_3^{-1}\sigma_2\sigma_1 P_{n+1,n+2})
            \delta(\tau_4^{-1}\tau_3^{-1}\tau_2\tau_1 P_{m+1,m+2})$}
\put (0,80){$+\lambda_{qp}
\delta(\sigma_4^{-1}\sigma_3^{-1}\sigma_2\sigma_1)
\delta(\tau_4^{-1}\tau_3^{-1}\tau_2\tau_1 P_{m+1,m+2})$}
\put (0,75){$+\lambda_{qp}
\delta(\sigma_4^{-1}\sigma_3^{-1}\sigma_2\sigma_1 P_{n+1,n+2})
\delta(\tau_4^{-1}\tau_3^{-1}\tau_2\tau_1)$}

\put (95,45){\line(1,0){20}}
\put (85,45){$\sigma_1,\!\tau_1$}
\put (116,45){$\sigma_2,\!\tau_2$}
\put (100,40){$(n,m)$}
\put (20,45){$\delta(\zeta\sigma_1\sigma_2^{-1})
          \delta(\xi\tau_1\tau_2^{-1})$}
\put (97,52){cut}

\put (94,28){$(n\! +\! 1,m\! +\! 1)$}
\put (95,25){\line(1,0){20}}
\put (85,25){$\sigma_1,\!\tau_1$}
\put (116,25){$\sigma_2,\!\tau_2$}
\put (20,25){$\delta(\sigma_1\zeta'\sigma_2^{-1})
               \delta(\tau_1\xi'\tau_2^{-1})$}
\put (97,15){cut}

\put (105,45){\line(0,1){1}}
\put (105,47){\line(0,1){1}}
\put (105,49){\line(0,1){1}}
\put (105,51){\line(0,1){1}}
\put (105,53){\line(0,1){1}}

\put (105,25){\line(0,-1){1}}
\put (105,23){\line(0,-1){1}}
\put (105,21){\line(0,-1){1}}
\put (105,19){\line(0,-1){1}}
\put (105,17){\line(0,-1){1}}
\end{picture}
\\
{\it Fig. 13: Feynman rules.}\\ \\ }
\\

\section{Summing over fold configurations and simple cases on a lattice}
So far we have concentrated on the contribution to the partition function
from a {\em fixed} fold configuration (locus of singular points) in the
target space. We saw that the contribution of all maps with the specified
fold configuration reduces to a combinatorial calculation for which we
had the Feynman rules of the previous section. In order to obtain the
full partition function we have to sum over all the fold configurations
as well. We do not know how to perform this task in general.
We also have to admit
 that in the Feynman rules of the previous chapter there
is no small parameter that limits the number of fold intersections.

In this section, however, we will demonstrate what is involved in the
(maybe) simpler problem of summing over fold configurations which separate
$(1,1)$ regions from $(0,0)$ regions in a toroidal target space. We assume
that no higher $(m,n)$ regions exist and thus the ``combinatorial''
contribution is trivial. In this case the geometrical picture is
that of Riemann surfaces of various genera smashed on a spherical
target space in a non-overlapping way. The problem is to sum over all such
configurations with the weight of
\begin{equation}
 \lmst^{\sum_i (2-2g_i)}e^{-2\mu\sum_i A_i}
\end{equation}

where the $A_i$-s are the areas of the bounded $(1,1)$ regions
and the $g_i$-s are the
genera of the smashed surfaces which correspond to the bounded regions.
We will formulate the problem on a lattice and map the lattice problem
to a perturbed Ising model with a triple interaction round a face.
This model is known as the {\em Baxter-Wu model}\cite{BAXWU}
\footnote{ We thank A. Zamolodchikov for pointing our attention to this
reference}.

It turns out to be convenient to work on a hexagonal lattice.
We will thus pick a hexagonal lattice $\Gamma$ for the target space, with
hexagons as faces. We put spin variables $\sigma_k = \pm 1$ on each face
and interpret a given configuration of $\{\sigma_k\}$ as a partition of the
target space into regions in such a way that the hexagons $k$
with $\sigma_k=+1$ belong to the $(1,1)$ regions and the hexagons with
$\sigma_k = -1$ belong to the $(0,0)$ regions.
The index $k$ is a vertex of the dual lattice $\Gamma^*$.
The total area is given by
$$
2 \sum_i A_i = \sum_{k\in {\Gamma^*}} (1 + \sigma_k)
$$
The Euler characteristic of the figure $\sum_i(2-2g_i)$ is given by Euler's
formula $V+F-E$ where we use the triangulation that is given by the lattice
$\Gamma$. For the number of faces $F$ we count the number of hexagons
in a $(1,1)$ region with weight 2. Note that hexagons in a $(0,0)$ region
do not contribute to $F$.
For the number of edges $E$ we count an edge twice if it is
{\em inside} a $(1,1)$ region, we count it once if it is on the boundary
between a $(1,1)$ and a $(0,0)$ region and we don't count it if it is in a
$(0,0)$ region. We will use the relation $E=3F$ (since every face has six
different edges and every edge is common to two faces. For the vertices $V$
we again count a vertex twice inside a $(1,1)$ region, once on the boundary
and don't count a vertex inside a $(0,0)$ region.
We can express these relations in terms of the spin variables $\{\sigma_k\}$
as follows:
\begin{eqnarray}
\sum_i (2-2g_i) &=& V+F-E  = V-2F                    \nonumber\\
F &=& N_{faces}+\sum_{k\in {\Gamma^*}} \sigma_k      \nonumber\\
V &=& \sum_{\Delta_{klm}\in {\Gamma^*}}
     (1-{1\over 4}(\sigma_k + \sigma_l + \sigma_m + \sigma_k\sigma_l\sigma_m))
              \nonumber\\
  &=& N_{vertices} - {3\over 2}\sum_{k\in {\Gamma^*}} \sigma_k
      -{1\over 4} \sum_{\Delta_{klm}\in {\Gamma^*}}\sigma_k\sigma_l\sigma_m
                                                     \nonumber
\end{eqnarray}

where $N_{faces}$ is the number of faces in the lattice $\Gamma$ and
$N_{vertices}$ is the number of vertices in $\Gamma$.
$\Delta_{klm}\in {\Gamma^*}$ is an elementary triangular cell in $\Gamma^*$
with vertices $k,l,m$.

Thus we obtain
\begin{eqnarray}
e^{-2\mu \sum_i A_i} \lmst ^{\sum_i (2-2g_i)} &=&
  e^{-(\mu + 2\log\lmst)N_{faces} +\log\lmst N_{vertices}}
 \nonumber\\
\cdot
 exp\{ -(\mu + {7\over 2}\log\lmst)\sum_{k\in {\Gamma^*}} \sigma_k &-&
{1\over 4}\log\lmst\sum_{\Delta_{klm}\in {\Gamma^*}}\sigma_k\sigma_l\sigma_m \}
\end{eqnarray}

Thus the Hamiltonian is given by
$$
{\cal H} = -(\mu + {7\over 2}\log\lmst)\sum_{k\in {\Gamma^*}} \sigma_k
-{1\over 4}\log\lmst\sum_{\Delta_{klm}\in {\Gamma^*}}\sigma_k\sigma_l\sigma_m
\label{BAXWUH}
$$
This is the Hamiltonian of the Baxter-Wu model in an external magnetic
field \cite{BAXWU}.

We will recall the results of \cite{BAXBOOK} and briefly describe their
consequences here:
\begin{enumerate}
\item
  The model with hamiltonian
$K\sum_{\Delta_{klm}\in {\Gamma^*}}\sigma_k\sigma_l\sigma_m$
has a second order phase transition for $K_c = \pm 0.4406\dots$ and zero
magnetic field.
In our case, we get
$$
\lmst = 0.171\dots,\qquad \mu=6.168\dots
$$
(Recall that $\mu$ is the cosmological constant per lattice cell.)

\item
Near the critical point let $\Delta K = K-K_c$.
Then, in the vicinity of $K_c$, the
 specific heat of the model behaves as $|\Delta K|^{-{2\over 3}}$.
The spontaneous magnetization for $K<K_c$ behaves as
$|\Delta K|^{1\over 12}$. It can be shown from scaling assumptions
\cite{BAXBOOK} that the partition function behaves as
$\log Z \sim |\Delta K|^{4\over 3} F(H |\Delta K|^{-{5\over 4}})$
where $H$ is a small magnetic field and $F$ is some function
(which is different for $\Delta K > 0$ and $\Delta K < 0$).
In our case, we replace $|\Delta K|$ by $\Delta\lmst$ and $H$ by
$\Delta\mu$ (since $H$ scales as $|\Delta\lmst|^{5\over 4}$ the
extra ${7\over 2}\log\lmst$ in (\ref{BAXWUH}) do not contribute to
the scaling behavior).
We get
$$
\log Z \sim |\Delta \lmst|^{4\over 3}
          F(\Delta\mu |\Delta \lmst|^{-{5\over 4}})
$$

\item
At the critical point itself the correlation functions
behave as $r^{-{1\over 4}}$.
Returning to the string model, we can deduce the tachyonic two-point
functions.
Denoting by  $Y$  a target space coordinate
we can write the operator that corresponds
to the tachyon as the Fourier transform of the cover number at $Y$ namely,
$T_k=\int d^2 Y e^{k\cdot Y} n(Y)$ where  $n(Y)= \int d^2\xi \delta
(Y-X(\xi)) d^2 \xi$.  A two-point function of tachyons can be
rewritten as
$\bra T_kT_{-k}\ket= \int d^2 Y e^{k\cdot Y} \bra n(Y)n(0)\ket$.
The restriction to unfolded maps makes $n(Y)$ independent of $Y$ and
therefore, all the two point functions (apart from $T_0$) vanish. In the
presence of folds  the cover numbers at different target space
points are correlated  and the tachyon correlators  are non-trivial.
In our case we get
$$
\bra n(Y)n(0)\ket = 1 + 2 \bra \sigma\ket + \bra \sigma(Y)\sigma(0)\ket
$$
and using  $\bra n(Y)n(0)\ket\sim |Y|^{-{1\over 4}}$ we get the tachyon
two-point function for $k\neq 0$
$$
\bra T_k T_{-k}\ket\sim |k|^{-{7\over 4}}
$$
\end{enumerate}

At this stage we do not know how to interpret this result in comparison
with    the corresponding correlators of string models.



\section{Interpretation of folds as particle trajectories}

The string theory that describes the large $N$ limit of $SU(N)$
2D Yang-Mills theory was developed in \cite{GT2}
(and following it in \cite{Mina} for $U(N)$).
it is a theory of maps with no folds.
The string coupling constant is identified with $1/N$,
the string tension is given by $\lambda/2$ where $\lambda$ is related to the
gauge coupling $\tilde{g}$ by $\lambda=\tilde{g}^2 N$.
 Maps with a branch
point in an infinitesimal area $dA$ have a weight of $\lambda dA$.
There is, however, a mysterious feature of this string theory which is the
$2-2G-B$, where  $G$ is the target space genus  and $B$ is the number of its
boundary components,
constant points in the target space.
\footnote{Those points have lately received a new
interpretation in \cite{Moore}.}

For these points the maps are allowed
to have an arbitrary branch point connecting arbitrary sheets and have
weight $1$ instead of an infinitesimal weight of $\lambda dA$.

We wish to incorporate folds and singular points in this framework.
Let's consider a sphere target space with a simple fold that divides
the target space into two simply connected areas $A$ and $\tsp-A$.
According to \cite{GT2} the natural operators which capture the geometry
at the boundary $\partial A$ are the ``generalized Frobenius characters''
which are given by equation (A.7) of \cite{GT2}
\begin{eqnarray}
&&\Upsilon_{\bar{\tau}\sigma}(U,\Udg)
=\sum_{v\subset T_\sigma}\sum_{v'\subset T_\tau,v\approx v'}
(-1)^{K_v} C_v\Upsilon_{\sigma\setminus v}(U)
     \Upsilon_{\tau\setminus v'}(\Udg) \nonumber\\
&=&\sum_{v_1,v_2,\dots}(-1)^{\sum v_i}\prod_i (i^{v_i} v_i!
\binom{\sigma_i}{v_i}\binom{\tau_i}{v_i})
\prod_i(Tr U^{\sigma_i-v_i})\prod_i(Tr (\Udg)^{\tau_i-v_i}) \nonumber\\
&&
\label{GENFROB}
\end{eqnarray}
where $\sigma$ and $\tau$ are permutations, $T_\sigma$ and $T_\tau$
 are the conjugacy classes
of $\sigma$ and $\tau$ respectively and
 $K_\sigma$ is the number of cycles of $\sigma$.
$C_v$ is given by
$$
C_v=\prod_i i^{v_i} v_i!
$$
where $v_i$ is the number of cycles in $v$ of length $i$, and
$$
\Upsilon_\sigma(U)=\prod_{i=1}^{K_\sigma}(Tr U^{k_i})
$$
where $k_i$ is the length of the $i$-th cycle of $\sigma$.

According to equation (4.8) of \cite{GT2}
the functional integral of the
Yang-Mills action on one side of the fold (say A) times
the value of $\Upsilon_{\bar{\tau}\sigma}(U,\Udg)$, where $U$ is the
holonomy of the gauge field around $\partial A$,
is given by the sum of maps
that cover $A$ so that the boundary is made up of $K_\sigma+K_\tau$
components. $K_\sigma$ components that are boundaries of sheets of positive
orientation and $K_\tau$ components that are boundaries of sheets of
the opposite orientation. The $i$-th component from the $K_\sigma$ has a
winding of $\sigma_i$ and the $j$-th component from the $K_\tau$ has a
a winding of $\tau_j$.
When there is no fold on $\partial A$
$A$ and $\tsp-A$ are glued by the gluing formula
(4.14) of \cite{GT2}
$$
\int dU\Upsilon_{\bar{\tau}\sigma}(U,\Udg)
       \Upsilon_{\bar{\tau}'\sigma'}(\Udg,U)
       =\delta_{T_\sigma,T_{\sigma'}}\delta_{T_{\tau'},T_\tau}C_\sigma C_\tau
$$
This means that the boundary of $A$ (which is parameterized by $T_\sigma$
and $T_\tau$) is glued to the boundary of $\tsp-A$ (which is
parameterized by $T_{\sigma'}$ and $T_{\tau'}$) so that each $k$-cycle
of $\sigma$ is attached to a $k$-cycle of $\sigma'$ and each $k$-cycle
of $\tau$ is attached to a $k$-cycle of $\tau'$.

We wish to find the matrix elements of a fold between the states
$\langle\sigma'\tau'|$ and $|\sigma\tau\rangle$. The relevant target space
is a cylinder
whose upper boundary corresponds to the permutation classes $T_{\sigma'}$
and $T_{\tau'}$ and whose lower boundary corresponds to
$T_\sigma$ and $T_\tau$. In the middle of the cylinder there is a fold
(see Fig.14).

\putfig{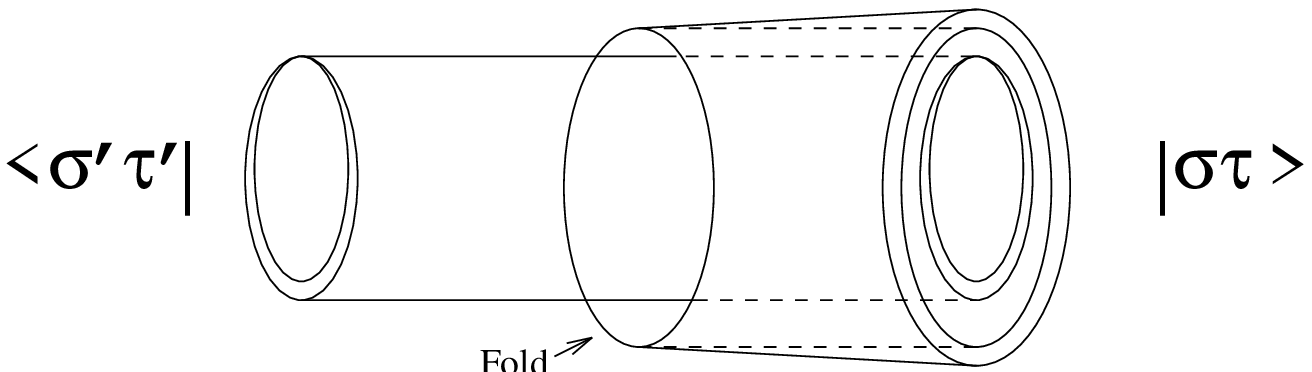}{Fig. 14: The matrix element of a fold on a
                       cylindrical target space.}{100mm}

If $\sigma\in S_n$, let $D_k\sigma\in S_{n+k}$ denote a permutation in
the conjugacy class that is obtained from the conjugacy class of $\sigma$
by adding one extra $k$-cycle. Similarly let $D_k^{-1}\sigma\in S_{n-k}$
denote a permutation in the conjugacy class that is obtained from the
conjugacy class of $\sigma$ by extracting one $k$-cycle.
A fold with no singular points can either add one $1$-cycle to $\sigma$ and
one $1$-cycle to $\tau$ or take away one $1$-cycle from $\sigma$ and one
$1$-cycle from $\tau$. Let $\opfd$ denote the operator of the fold.
Then
$$
\opfd|\sigma,\tau \rangle = |D_1\sigma,D_1\tau \rangle
                   + \sigma_1\tau_1 |D_1^{-1}\sigma,D_1^{-1}\tau \rangle
$$
where
the $\sigma_1\tau_1$ coefficient in the second term is because we can
take out any one of the $\sigma_1$ $1$-cycles of $\sigma$ and any one of the
$\tau_1$ $1$-cycles of $\tau$.

{}From (\ref{GENFROB}) we see that
\begin{eqnarray}
Tr(U)Tr(\Udg)\Upsilon_{\bar{\tau}\sigma}(U,\Udg)
&=&\Upsilon_{\overline{D_1\tau}D_1\sigma}(U,\Udg)
+\sigma_1\tau_1
\Upsilon_{\overline{D_1^{-1}\tau}D_1^{-1}\sigma}(U,\Udg) \nonumber\\
&+&(\sigma_1+\tau_1+1)\Upsilon_{\overline{\tau}\sigma}(U,\Udg)
\label{UUDGOP}
\end{eqnarray}
where we have used the combinatorial identity
\begin{eqnarray*}
\lefteqn{
\sum_v (-1)^v\binom{a}{v}\binom{b}{v}v! x^{a+1-v}y^{b+1-v} =}\\
&&(a+b+1)\sum_v (-1)^v\binom{a}{v}\binom{b}{v}v! x^{a-v}y^{b-v} \\
&+&ab\sum_v (-1)^v\binom{a-1}{v}\binom{b-1}{v}v! x^{a-1-v}y^{b-1-v} \\
&+&\sum_v (-1)^v\binom{a+1}{v}\binom{b+1}{v}v! x^{a+1-v}y^{b+1-v}
\label{ComId}\end{eqnarray*}
for $x=Tr(U)$, $y=Tr(\Udg)$, $a=\sigma_1$ and $b=\tau_1$.
Equation (\ref{UUDGOP}) expresses the fact that the two boundaries of
opposite orientations that are created by the operator $Tr(U)Tr(\Udg)$,
can be formed by either a fold in the upward direction (the first term
in the rhs), a fold that goes downward (the second term),
 or without a fold -- by cutting a sheet into two (the last term).
We get the symbolic relation
$$
\opfd = (Tr(U)Tr(\Udg)-1)-(\sigma_1+\tau_1)
\label{OPFDUU}
$$
If it were only the $Tr(U)Tr(\Udg)-1$ term, we could understand a fold
as a trajectory of a particle in the adjoint representation of $SU(N)$
(or $U(N)$) since it would carry the weight
$$
Tr(U) Tr(\Udg)-1 = Tr_{Adj} (P e^{i\oint A^\alpha dx_\alpha})
$$
The second term $\sigma_1+\tau_1$ means that we have to add a particle
which is a singlet
  ghost (because of the $(-)$ sign) and  has a peculiar
interaction with the gauge field such that it has a multiplicity of
$\sigma_1+\tau_1$.
At this stage the nature of
 this apparent ghost is not clear to us.

Next, we consider a fold on the cylinder with one singular point on it.
Suppose that there are two more sheets {\em above} the fold.
Denote by $(n,m)$ the numbers of sheets below the fold,
whose boundary components are parameterized by $\sigma$ and $\tau$.
We can pick some labeling of the $(n+m)$ sheets at the singular point $P$
on the fold, and then $\sigma$ and $\tau$ will not be just conjugacy classes
but will be unique permutations that specify how the sheets change after
a permutation starting at $P$.
We will work with the following settings:
To the left of $P$ sheet `$a$' is connected
to the fold and sheet `$b$' goes over to sheet `$x$' below the fold.
To the right of $P$ sheet `$b$' is
connected to the fold and sheet `$a$' goes over to
sheet `$x$'. If `$x$' is one of the $n$ orientability $(+)$ sheets, then
the fold changes $|\bar{\tau}\sigma\rangle$ into
$|\bar{\tau}\rho_{x,n+1}\sigma\rangle$ where $\rho_{x,n+1}\in S_{n+1}$ is the
permutation that switches the two elements $x$ and $n+1$.
Similarly, when $x$ is one of the $m$ orientability $(-)$ sheets,
the fold changes $|\bar{\tau}\sigma\rangle$ into
$|\overline{\rho_{x,m+1}\tau}\sigma\rangle$.
So we can write symbolically the operator
that corresponds to the singular point $P$ on
a fold that is facing downwards as:
$$
\opslr = \sum_x\rho_{x,n+1}+\sum_y\rho_{y,m+1}
$$
but
$$
\rho_{x,n+1}=\sum_{x<y\leq n+1} \rho_{x,y}
            -\sum_{x<y\leq n} \rho_{x,y}
$$
The operator $\sum_{x<y\leq n} \rho_{x,y}$ is related to the operator
$Tr(F(P_-)^2)$ where $F(P)$ is the YM field strength at the point $P$
and the subscript $P_-$ indicates that we actually have to take the field
strength a little bit below the point $P$ (that is, a little bit below the
fold).
To see the relation we note that if there was no fold, the string operator
$\sum_{x<y\leq n} \rho_{x,y}$ changes a boundary state of
$|\bar{\tau}\sigma\rangle$ as follows
$$
|\bar{\tau}\sigma\rangle \rightarrow
\sum_{x<y\leq n}|\overline{\rho_{x,y}\tau}\sigma\rangle
$$
Thus its matrix elements are given precisely by a sum of maps that have
a branch point at $P$ (and the weights of the maps are according to the
rules of \cite{GT2}).
By differentiating
the equation for the partition function of $YM_2$, u
written in terms of maps
(equation (3.6) of \cite{GT2}) with respect to $\lambda$
we obtain the string analog of the operator insertion of $Tr(F^2)$ at a
point $P$:
$$
{N\over{4\lambda^2}}Tr(F^2)=
-\hlf(n+m)+{{(n-m)^2}\over{2N^2}}
-{1\over  N}(\sum_{x<y\leq n} \rho_{x,y}
           +\sum_{\bar{x}<\bar{y}\leq m} \rho_{\bar{x},\bar{y}})
\label{STRINGF2}
$$
The last term has the following meaning: when it is inserted in expectation
values (of Wilson loops) it counts the maps with a branch point at $P$.
It arises from the contribution of $\lambda$ to the weight of maps with
branch points at $P$. $n(P)$ and $m(P)$ are the number of covers of $(+)$
and $(-)$ orientability at $P$.
To get (\ref{STRINGF2}) we summed over connecting tubes and contracted handles
at $P$.
Thus we see that
$$
\sum_{x<y\leq n} \rho_{x,y}
           +\sum_{\bar{x}<\bar{y}\leq m} \rho_{\bar{x},\bar{y}}
=
-{{N^2}\over{4\lambda^2}}Tr(F^2)+N\hlf(n+m)-{{(n-m)^2}\over{2N}}
$$

We thus  get the  following expression for the operator of a singular point
on a fold
$$
\opslr = 2N-{{N^2}\over{4\lambda^2}}(Tr(F(P_+)^2)-Tr(F(P_-)^2)
\label{OPpof}
$$
where $P_+$ means a little bit above the fold and $P_-$ means a little
bit below the fold.

Suppose the operator $\opfd$ of a fold with no singular points
was just $Tr(U)Tr(\Udg)-1$, which is the
trace in the adjoint representation of the Wilson loop along the fold.
This would mean, as mentioned above, that the interaction of a fold with
the gauge field is exactly like that of
a particle in the adjoint representation
of an $SU(N)$ gauge group.
In that case  the operator $\opslr$ would give a mass insertion of
$2N$ and a peculiar interaction with the jump in $Tr(F^2)$ along the fold.


Another QCD operator that
may be related to the operator $\opslr$  is an insertion of
$g_{\mu\nu} D_\tau F^{\tau\nu}$. This operator is explicitly not APD
invariant and is  inserted  into Wilson loops as follows
$$
Tr_{Adj}Pe^{\oint A_\mu dx^\mu} \rightarrow
Tr_{Adj}Pe^{\oint (A_\mu+\rho g_{\mu\nu} D_\tau F^{\tau\nu}) dx^\mu}
\label{WiLi}
$$
We can interpret this insertion pictorially.
Consider a Wilson loop in the adjoint. As mentioned above, according to
\cite{GT2} there are three groups of string maps that are to be considered.
Those that have two more sheets to one side of the Wilson loop, those that
have two more sheets to the other side, and those that have the same number
of sheets on both sides.

An insertion of $g_{\mu\nu} D_\tau F^{\tau\nu}$ at a point $x$ can be
interpreted as deforming one of the two boundaries of the Wilson loop at the
point $x$ slightly outwards (say by a square) and then subtracting a similar
deformation slightly inwards, and taking the $O(\Delta A)$ coefficient
\cite{KAZAKOV}. In the string picture the maps that contribute to the
$O(\Delta A)$ are those that have a branch point in the $\Delta A$ region,
and there is also a contribution from the $e^{\lambda \Delta A}$ term for
maps which do not necessarily have a branch point in the $\Delta A$ region.
The problem is that we get a contribution with a negative sign for the
maps that have two more sheets on the other side!

We remark, that the $\Omega$-points of \cite{GT2} do not
cause a real difficulty since every boundary component creates one
$\Omega$-point on one side of it and an $\Omega^{-1}$-point on the other
side. Thus, a fold that has one boundary from every orientation creates
a pair of an $\Omega$-point and an $\Omega^{-1}$-point
on each side of it, and the pairs cancel.

We will end by summarizing the discrepancies that are still left in the
interpretation of folds as particle trajectories.
\begin{itemize}
\item
The correspondence between
 maps with a fold boundary and maps that calculate the
expectation value of a Wilson loop in the adjoint representation is not
complete because the latter include maps that are described by cutting a smooth
sheet along a trajectory (thus creating a two boundaries of opposite
orientations) and not just by maps that are obtained by cutting along a fold.
The extra maps are described by the $\sigma_1+\tau_1$ term of (\ref{OPFDUU}).

\item
The peculiar interaction of the fold with the discontinuity of $Tr(F^2)$
along it does not have a satisfactory Lagrangian description.

\item
The quartic vertices of the ``Feynman diagrams'' of the previous section
were not explained in terms of a lagrangian.
\end{itemize}

\section{ Summary and Discussion}
In this paper we have explored the role of
folded maps in  two dimensional string
theory.  We have removed several obstacles on the way toward
computing  the contribution
of folds to the string  partition function.
In this respect we have made a certain
progress in the program of quantizing the 2D NG theory.
 Folds have been shown to admit
a behavior of particles with special interaction with non-abelian gauge fields.
This may turn out to be relevant in the search of stringy nature of $QCD_2$.

The precise relation between the NG theory and the Polyakov
string theory at the
quantum level has not yet been fully clarified. In 2D one expects that the
non-critical string theory of $c=1$ is related to the NG theory\cite{Gross}.
This correspondence faces several problems, for instance  what Liouville
theory  corresponds to  higher genus 2D target space.

To the extent that the correspondence contains certain truth,
one would associate
the NG folds with the massless tachyons of the Liouville theory.
Indeed  our
analysis  partially supports such an interpretation.
Denoting by  $Y$  a target space c-number, we can write the operator that
corresponds
to the tachyon as the Fourier transform of the cover number at $Y$ namely,
$T_k=\int d^2 Y e^{k\cdot Y} n(Y)$ where  $n(Y)= \int d^2\xi \delta
(Y-X(\xi)) d^2 \xi$.  A two-point function of tachyons can be
rewritten as $\bra T_kT_{-k}\ket= \int d^2 Y e^{k\cdot Y} \bra n(Y)n(0)\ket$.
The restriction to unfolded maps makes $n(Y)$ independent of $Y$ and
therefore, all the two point functions (apart from $T_0$) vanish. In the
presence of folds  the cover numbers at different target space
points are correlated  and the tachyon correlators  are non-trivial.
 It is easy to realize that tachyons indeed have
vanishing $det J$.
It seems that they are associated with a fold-``antifold" configuration.

In the present work we have put forward the idea
 that folds should be associated with
particle trajectories. At present we have certain evidence that
support it,  even though, a full understanding of the nature of those
particles is still ahead of us.  The contribution to the partition
function  of  a single fold
enclosing a region covered with three sheets resembles the contribution of
a particle with mass $\lambda_{st}\rho$. Moreover, an analysis
based on the  methods of \cite{GT2} indicates that
indeed a fold could
be understood as  a  trajectory of a particle in
 the adjoint representation with certain peculiar interaction with the gauge
fields.
So far this  peculiar interaction of the fold with the discontinuity of
$Tr(F^2)$ along it does not have a satisfactory Lagrangian description.
A full correspondence between
   maps with a fold boundary and maps that calculate the
expectation value of a Wilson loop in the adjoint representation was not
established. For the  case of a Wilson loop  the maps  are described
by cutting a smooth sheet along a trajectory (thus creating a two boundaries of
opposite orientations) and not just by maps that are obtained by cutting along
a
fold. The extra maps are described by the $\sigma_1+\tau_1$ term of
(\ref{OPFDUU}). Clearly, this important issue deserves more study.

The stringy picture of $YM_2$ is based on the perturbative ${1\over N}$
expansion. The weighted sum over unfolded maps was constructed to reproduce
this expansion. So far we lack a stringy action which encodes this
sum over maps. Recently, some progress has been made toward this goal.
First, it was argued \cite{Mina} that to account for the $U(N)$ (rather
than $SU(N)$) partition function, the only point singularities which appear
in the maps are branch points and the so called ``$\Omega$-points''.
One has to insert $2G-2$ of the latter and, therefore, for a target space
with topology of a torus ($G=1$), only branch points are left.
In this case, summing over holomorphic maps from a torus world sheet reproduces
the chiral $YM_2$ partition function\cite{BCOV}. The sum over holomorphic
maps for a toroidal target space is encoded in the topological $A$ type
sigma model\cite{WTSM}.
A proposal following these lines was made in
\cite{Horava} where the topological string theory is based on
harmonic maps.
Very recently \cite{Moore}
 a modified topological string theory was introduced. In that
formulation a new interpretation of the $\Omega^{-1}$ points was given.
Moreover, many
aspects of the large $N$ limit of the $SU(N)$ $YM_2$
 theory were reproduced in the zero coupling limit.

Another idea concerning
 the string action\cite{GT1} was that it would be a NG action
with an additional  fold suppressing term.
When we started this work, our hope was  to convert the full
NG action into a target space action  that has two parts, a $YM_2$ term and an
action
of interacting particles. In that case we might be able to add additional terms
to cancel the contribution of the folds.
This would provide the desired $YM_2$ string action.
Alternatively, a world sheet theory describing
the folds could be amended so  that the contribution of folds to the generating
functional would have been cancelled. At present, since we do not have a full
understanding of folds in neither of those  two pictures, we cannot
use our approach to get information pertaining to the string action.

Another model that may shed light on this problem is the one dimensional
analogous
model.  In the space of maps from a world-line to a one dimensional target
space,
singular points  are the analogs of the line singularities  in the 2D case.
The induced measure, the resulting target space quantum mechanical theory and
additional related topics are   under current  investigation.

The construction of the stringy action of $YM_2$ and $QCD_2$ is still
a challenging and important problem.

As was explained in section (2) the contribution of maps with singular points
are characterized
by a set of different parameters associated with the various types of
singularities.
Thus, it is implied that the NG action is not one string model but
it corresponds to a family of  string models.
In the stringy YM theory one is forced to set those parameters to be equal.
It might be that  the
parameters are determined by the requirements of unitarity of
the string theory and by the requirement that under a renormalization
group flow to the infra-red the parameters remain relevant.
In fact there is
such an ambiguity in any infinite dimensional measure.
In $\prod_{k=1}^n dx_k$ with metric $\sum_k dx_k^2$ we can rescale the
measure by $\lambda$ and this would lead to rescaling of the metric by
$\lambda^{2/n}$. When $n\rightarrow\infty$ we can rescale the measure by a
finite amount without changing the metric at all!
It is interesting to wonder whether such a type of
``rescaling'' can remove the
crumpling phase transition of the $D=2$ string.
Namely, that the crumpling transition happens when maps with folds get
counted with a weight that goes to infinity when the cut off goes to
infinity. This is natural from the point of view of lattice regularization
since the folds can pass anywhere on cell boundaries, and the number of maps
with folds is proportional to the number of paths through cell boundaries
which goes to infinity as the lattice constant goes to zero. However,
when we damp the weight of the folds and make it finite (a finite measure
of random walks in the target space) by adjusting the relative coefficient
between the measure of maps with folds and maps without folds, maybe we can
get rid of the crumpling transition.

Renormalizability and unitarity are two very important questions that
were  not addressed in our work.
Even though we do not encounter  divergences in the computations
that we have performed. It is plausible that the contributions from
infinitely small folds do diverge. In this case
renormalizability arguments probably force us to include additional special
points like ``spiral-points''.
In section 4 we have hinted about the possibility that due to
this type of renormalizability configurations with branch points ``close" to
folds may have a significant weight.
A detailed analysis of these issues is under current investigation.

In the present work only maps from compact world sheets were considered.
We have thus analyzed folds only in the framework of closed string theory.
A discussion of folds in open strings  both in the classical and semi-classical
was presented in \cite{Bars}. It was found that the string  normal modes
correspond to a completely folded strings having equal lengths that are
oscillating in that form. The exact interpretation of  those folds in terms
of the $QCD_2$ picture is still unclear. The relations between that canonical
analysis and our path-integral formulation has to be further investigated.

Another direction that has to be further explored is the space of maps to a
higher genus target space.  It seems that the latter is
  much more complex then the space of maps to the sphere.

As is evident from the present work the summation over folds is difficult.
It may be advantageous to try and sum over specific families of folds.
This was done in section (5), where the simple case of a non-self-overlapping
string was investigated. In this case the summation problem was mapped into
a lattice model which turned out to be the Baxter-Wu three-spin model in an
external magnetic field. It would be important to find other lattice or
continuum models which sum more complicated families of folds. This direction
is under current investigation. A word of caution; even if we manage to sum
over certain families of folds, there is apriori no guarantee that this
partial sum will capture the physics of the full problem.

While completing this work we received two manuscripts that discuss related
topics\cite{FrPa}.

\vspace{1.5cm}
\centerline{\bf Acknowledgment}
\vspace{0.5cm}
 We would like to thank P. Di. Francesco, D. Kutasov, H. Neuberger
 and  N. Seiberg
 for useful conversations.  We would
especially like to thanks W. Taylor and A. Zamolodchikov for helpful
discussions and for
their enlightening
comments on various parts of this work.

\newpage
\section*{Appendix A : Calculation of the fold matrix}
We will calculate the fold matrix for a fold that separates a region
of cover-numbers $(n,m)$ from a region of
cover-numbers $(n+1,m+1)$. We will assume that there is
no branch point exactly on the fold.
This is certainly true in the ``semi-classical'' case as explained previously.
However, we will consider spiral points (as in Figs.7,9) as well.

To check whether a specific point $P$ on a fold
is a branch point or not, we have to
surround the point by a small loop $\beta$ in the target-space. Each point
in the loop corresponds to more than one point on the world-sheet if the
number of covers at the vicinity of $P$ is greater than one. Next, we have to
traverse the loop in the {\em world-sheet}, that is, pick a starting point
on the world-sheet and go along $\beta$ on the corresponding sheets.
In our route we will automatically switch sheets, and when we run into a
fold we will have to reverse direction and return backwards in $\beta$ but
on a reversely oriented sheet. Eventually we will return to our starting
point {\em on the same sheet}. Counting our total winding number will give
us the order of the branch point. If it is greater than 1 then $P$ is a
branch-point, if the winding number is 1 then the point is
a simple singular point on the fold, and if
the winding number is zero then the point is {\em inside} a fold.

An example for the case of the smashed handle of Fig.3 will explain
this procedure.
We check the point P which is the common endpoint of $\gamma_1$ and
$\gamma_2$ of (Fig.3). To the right of $\gamma_1,\gamma_2$ there is only one
sheet which we denote `$d$'.
To the left of $\gamma_1,\gamma_2$ there are three
sheets, two of them $(a,c)$ with the same orientation as `$d$' and the third
($b$) with the opposite orientation. At the vicinity of $\gamma_1$ `$d$'
continues to `$c$' and `$a$'
is joined to `$b$' by a fold, whereas near $\gamma_2$
`$d$' continues to `$a$' and `$c$' is
joined to `$b$' by a fold. The loop $\beta$
begins on `$d$', when it passes over $\gamma_2$ it continues on `$a$'.
$\beta$ changes direction when it meets $\gamma_1$
and returns on `$b$'. On $\gamma_2$ there is a second
change of direction after which $\beta$ continues on `$c$'
and at last, it returns to the starting point when `$c$' meets `$d$' under
$\gamma_1$. The total winding number is $1$ so that, as expected, this is no
branch-point.

We turn to the more general case. Assume there are $n$ sheets of
orientation $+1$ and $m$ sheets of orientation $-1$ to the right of
$\gamma_1,\gamma_2$, and there is one sheet more of each orientation to the
left of $\gamma_1,\gamma_2$.
Denote the sheets to the right as $\{a_i\}_{i=1}^{n}$ and
$\{b_j\}_{j=1}^m$ and the sheets to the left as $\{c_i\}_{i=1}^{n+1}$ and
$\{d_j\}_{j=1}^{m+1}$.
 We will first check the case that the folds $\gamma_1$ and $\gamma_2$ do
not have a sheet in common. Then, we can label the sheets so that  along
$\gamma_1$ -- $c_1$ and $d_1$ are joined by the fold and so that $b_j$ is the
continuation of $d_{j+1}$ and $a_i$ is the continuation of $c_{i+1}$.
Along $\gamma_2$ we assume that $c_{n+1}$ and $d_{m+1}$ are joined by the
fold, and that $a_i$ is the continuation of $c_{\sigma(i)}$ for some
$\sigma\in S_n$ and likewise $b_j$ is the continuation of $d_{\tau(j)}$
for some $\tau\in S_m$.

\parbox[c]{140mm}{
\begin{picture}(140,100)

\thicklines
\put(5,50){\line(1,0){130}}

\put(63,48){\line(1,1){4}}
\put(63,52){\line(1,-1){4}}

\put(64,53){$P$}
\put(5,53){$\gamma_1$}
\put(132,53){$\gamma_2$}

\thinlines

\put(4,35){\line(1,5){3}}
\put(3,32){$d_1$}
\put(10,35){\line(-1,5){3}}
\put(9,32){$c_1$}

\put(15,35){\line(0,1){30}}
\put(14,32){$d_2$}
\put(14,67){$b_1$}
\put(20,35){\line(0,1){30}}
\put(19,32){$d_3$}
\put(19,67){$b_2$}
\put(23,40){$\cdots$}
\put(23,60){$\cdots$}
\put(30,35){\line(0,1){30}}
\put(29,32){$d_{m+1}$}
\put(29,67){$b_m$}

\put(40,35){\line(0,1){30}}
\put(39,32){$c_2$}
\put(39,67){$a_1$}
\put(45,35){\line(0,1){30}}
\put(44,32){$c_3$}
\put(44,67){$a_2$}
\put(48,40){$\cdots$}
\put(48,60){$\cdots$}
\put(55,35){\line(0,1){30}}
\put(54,32){$c_{n+1}$}
\put(54,67){$a_n$}

\put(69,35){\line(1,5){3}}
\put(64,32){$d_{m+1}$}
\put(75,35){\line(-1,5){3}}
\put(74,32){$c_{n+1}$}

\put(83,35){\line(0,1){30}}
\put(82,32){$d_{\tau(1)}$}
\put(82,67){$b_1$}
\put(91,35){\line(0,1){30}}
\put(90,32){$d_{\tau(2)}$}
\put(90,67){$b_2$}
\put(94,40){$\cdots$}
\put(94,60){$\cdots$}
\put(101,35){\line(0,1){30}}
\put(100,32){$d_{\tau(m)}$}
\put(100,67){$b_m$}

\put(113,35){\line(0,1){30}}
\put(112,32){$c_{\sigma(1)}$}
\put(112,67){$a_1$}
\put(121,35){\line(0,1){30}}
\put(120,32){$c_{\sigma(2)}$}
\put(120,67){$a_2$}
\put(122,40){$\cdots$}
\put(122,60){$\cdots$}
\put(130,35){\line(0,1){30}}
\put(129,32){$c_{\sigma(n)}$}
\put(129,67){$a_n$}

\end{picture}
\\
{\it Fig. 15: Gluing of the sheets on two sides of
        a singular point $P$ on a fold.
      The fold does not have a sheet in common for both sides of $P$.}\\ \\ }
\\

 Start now with $a_i$ and make $r_1$ windings until either
\begin{equation}
\sigma(\dots(\sigma(\sigma(i)-1)-1)\dots) - 1 = 1
\end{equation}
in which case we have a winding of $r_1$ and we finish, or else:
\begin{equation}
\sigma(\dots(\sigma(\sigma(i)-1)-1)\dots) = 1
\end{equation}
then we change to $d_1$ and make $r_2$ anti-windings until
\begin{equation}
\tau^{-1}(\dots(\tau^{-1}(\tau^{-1}(1)+1)+1)\dots) + 1= m+1
\end{equation}
then, we switch to $c_{n+1}$  and next to $a_n$ and make
$r_3$ more windings until:
\begin{equation}
\sigma(\dots(\sigma(\sigma(n)-1)-1)\dots) - 1 = i
\end{equation}
(we have to reach $a_i$ before we reach $c_1$ in order not to
get trapped in circles on a trajectory that does not contain $a_i$!)

So, to summarize, there are three possibilities of trajectories:
\begin{enumerate}
\item
  trajectories of the form
  \begin{equation}
  a_{t_1}\rightarrow c_{\sigma(t_1)} \rightarrow a_{t_2}
    \rightarrow c_{\sigma(t_2)}\rightarrow\dots\rightarrow a_{t_r}
  \end{equation}
  with
  \begin{eqnarray}
  t_{i+1} &=& \sigma(t_i)-1,\qquad i=1,\dots r-1 \\
  t_r     &=& t_1
  \end{eqnarray}
  the winding number is $r-1$ so in order not to have a branch point
   we require that this will not be possible for $r>2$.
\item
  trajectories of the form
  \begin{equation}
  b_{s_1}\rightarrow d_{\tau(s_1)} \rightarrow b_{s_2}
    \rightarrow d_{\tau(s_2)}\rightarrow\dots\rightarrow b_{s_r}
  \end{equation}
  with
  \begin{eqnarray}
  s_{i+1} &=& \tau(s_i)-1,\qquad i=1,\dots r-1 \\
  s_r     &=& s_1
  \end{eqnarray}
  the winding number is again $r-1$ so in order not to have a branch point
  we require that this will not be possible for $r>2$.
\item
  trajectories of the form
  \begin{eqnarray}
  d_1 &=& d_{s_1}\rightarrow b_{\tau^{-1}(s_1)} \rightarrow d_{s_2}
\rightarrow b_{\tau^{-1}(s_2)}\rightarrow\dots\rightarrow d_{s_r} \nonumber\\
      &=& d_{m+1}\rightarrow c_{n+1}\rightarrow a_n \nonumber\\
      &=& a_{t_1}\rightarrow c_{\sigma(t_1)}
                 \rightarrow\dots\rightarrow c_{\sigma(t_p)}=c_1
  \end{eqnarray}
  with
  \begin{eqnarray}
  s_1     &=& 1 \nonumber\\
  s_{i+1} &=& \tau^{-1}(s_i)+1,\qquad i=1,\dots,r-1 \nonumber\\
  s_r     &=& m+1 \nonumber\\
  t_1     &=& n \nonumber\\
  t_{i+1} &=& \sigma(t_i)-1,\qquad i=1,\dots,p-1 \nonumber\\
  \sigma(t_p) &=& 1
  \end{eqnarray}
  The winding number is  $r-p-1$ so in order to have no branch points
  we require that this will not be possible for $|r-p-1|>2$
  In fact, for $r-p=1$ this construction describes a spiral point of order
  $r$ as in Fig.9. For $r-p=2$ We get a spiral point of order $r-\hlf$
  as in Fig.7.

\end{enumerate}

 Next, for the case that the folds do have a sheet in common, we
 can label the sheets so that  along $\gamma_1$ -- $c_{n+1}$ and $d_1$ are
joined by the fold and that $b_j$ is the continuation of $d_{j+1}$ and $a_i$
is the continuation of $c_i$. Along $\gamma_2$ we assume that $c_{n+1}$
and $d_{m+1}$ are joined by the fold, and that $a_i$ is the continuation of
$c_{\sigma(i)}$ for some $\sigma\in S_n$ and likewise $b_j$ is the
continuation of $d_{\tau(j)}$ for some $\tau\in S_m$.

\parbox[c]{140mm}{
\begin{picture}(140,100)

\thicklines
\put(5,50){\line(1,0){130}}

\put(63,48){\line(1,1){4}}
\put(63,52){\line(1,-1){4}}

\put(64,53){$P$}
\put(5,53){$\gamma_1$}
\put(132,53){$\gamma_2$}

\thinlines

\put(4,35){\line(1,5){3}}
\put(1,32){$c_{n+1}$}
\put(10,35){\line(-1,5){3}}
\put(9,32){$d_1$}

\put(15,35){\line(0,1){30}}
\put(14,32){$d_2$}
\put(14,67){$b_1$}
\put(20,35){\line(0,1){30}}
\put(19,32){$d_3$}
\put(19,67){$b_2$}
\put(23,40){$\cdots$}
\put(23,60){$\cdots$}
\put(30,35){\line(0,1){30}}
\put(29,32){$d_{m+1}$}
\put(29,67){$b_m$}

\put(40,35){\line(0,1){30}}
\put(39,32){$c_1$}
\put(39,67){$a_1$}
\put(45,35){\line(0,1){30}}
\put(44,32){$c_2$}
\put(44,67){$a_2$}
\put(48,40){$\cdots$}
\put(48,60){$\cdots$}
\put(55,35){\line(0,1){30}}
\put(54,32){$c_n$}
\put(54,67){$a_n$}

\put(69,35){\line(1,5){3}}
\put(64,32){$d_{m+1}$}
\put(75,35){\line(-1,5){3}}
\put(74,32){$c_{n+1}$}

\put(83,35){\line(0,1){30}}
\put(82,32){$d_{\tau(1)}$}
\put(82,67){$b_1$}
\put(91,35){\line(0,1){30}}
\put(90,32){$d_{\tau(2)}$}
\put(90,67){$b_2$}
\put(94,40){$\cdots$}
\put(94,60){$\cdots$}
\put(101,35){\line(0,1){30}}
\put(100,32){$d_{\tau(m)}$}
\put(100,67){$b_m$}

\put(113,35){\line(0,1){30}}
\put(112,32){$c_{\sigma(1)}$}
\put(112,67){$a_1$}
\put(121,35){\line(0,1){30}}
\put(120,32){$c_{\sigma(2)}$}
\put(120,67){$a_2$}
\put(122,40){$\cdots$}
\put(122,60){$\cdots$}
\put(130,35){\line(0,1){30}}
\put(129,32){$c_{\sigma(n)}$}
\put(129,67){$a_n$}

\end{picture}
\\
{\it Fig. 16: Gluing of the sheets on two sides of
        a singular point $P$ on a fold. The fold has a sheet in common
        for both sides of $P$.}\\ \\ }
\\

We have the following restrictions:
\begin{enumerate}
\item
  from trajectories of the form
  \begin{equation}
  a_{t_1}\rightarrow c_{\sigma(t_1)} \rightarrow a_{t_2}
    \rightarrow c_{\sigma(t_2)}\rightarrow\dots\rightarrow a_{t_r}
  \end{equation}
  with
  \begin{eqnarray}
  t_{i+1} &=& \sigma(t_i),\qquad i=1,\dots r-1 \\
  t_r     &=& t_1  \label{RECUR1}
  \end{eqnarray}
  the winding number is $r-1$ so in order to have no branch points we require
  that this will not be possible for $r>2$
\item
  trajectories of the form
  \begin{equation}
  b_{s_1}\rightarrow d_{\tau(s_1)} \rightarrow b_{s_2}
    \rightarrow d_{\tau(s_2)}\rightarrow\dots\rightarrow b_{s_r}
  \end{equation}
  with
  \begin{eqnarray}
  s_{i+1} &=& \tau(s_i)-1,\qquad i=1,\dots r-1 \\
  s_r     &=& s_1  \label{RECUR2}
  \end{eqnarray}
  the winding number is again $r-1$ so in order to have no branch points
  we require that this will not be possible for $r>2$
\item
  trajectories of the form
  \begin{eqnarray}
  d_1 &=& d_{t_1}\rightarrow b_{\tau^{-1}(t_1)} \rightarrow d_{t_2}
\rightarrow b_{\tau^{-1}(t_2)}\rightarrow\dots\rightarrow d_{t_r}\nonumber\\
      &=& d_{m+1}\rightarrow c_{n+1}\rightarrow d_1
  \end{eqnarray}
  with
  \begin{eqnarray}
  t_1     &=& 1 \nonumber\\
  t_{i+1} &=& \tau^{-1}(t_i)+1,\qquad i=1,\dots,r-1 \nonumber\\
  t_r     &=& m+1 \label{RECUR3}
  \end{eqnarray}
  the winding number is  $r-1$ so in order to have no branch points
  we require that this will not be possible for $r>2$
\end{enumerate}

The first two cases correspond to a branch point that is on a ``spectator''
sheets (i.e. disjoint from the sheets of the fold in a small neighborhood
of $P$). The third case corresponds to a branch point that ``sits''
on the fold itself and can be described as a limiting case of a branch
point that approaches a simple singular point on a fold.

Returning to the fold matrix $F^{\sigma_1\tau_1}_{\sigma_2\tau_2}$,
it can be seen,
from all that has been said above,
that for a spiral point of the type of Fig.9 and of order $k$,
$\sigma_2$ and $\tau_2$ can be
written in terms of $\sigma_1$ and $\tau_1$ as:
\begin{eqnarray}
\sigma_2 &=& \sigma_1\circ((n+1)\, z_1 z_2\dots z_k)
\nonumber\\
\tau_2   &=& \tau_1\circ((m+1)\, w_1 w_2\dots w_k)
\end{eqnarray}
where $(a_1 a_2 \dots a_r)$ denotes a cyclic permutation in which
$a_1$ goes to $a_2$ , $a_2$ goes to $a_3$ and so on, and
$z_1,z_2,\dots,z_k,w_1,w_2,\dots,w_k$ are some sheet numbers.

For a spiral point of the type of Fig.7 we get:
\begin{eqnarray}
\sigma_2 &=& \sigma_1\circ((n+1)\, z_1 z_2\dots z_{k-1})
\nonumber\\
\tau_2   &=& \tau_1\circ((m+1)\, w_1 w_2\dots w_k)
\end{eqnarray}
or
\begin{eqnarray}
\sigma_2 &=& \sigma_1\circ((n+1)\, z_1 z_2\dots z_k)
\nonumber\\
\tau_2   &=& \tau_1\circ((m+1)\, w_1 w_2\dots w_{k-1})
\end{eqnarray}

If we include the ``spiral coupling constant'' $\rho_{sp}$ for points
of the type of Fig.9 and $\rho_{sp}'$ for points of the type of Fig.7,
we get:
\begin{eqnarray}
F^{\sigma_1\tau_1}_{\sigma_2\tau_2}&=&
 \sum_{k=1}^\infty \#\{z_1,z_2,\dots,z_k\;,w_1,w_2,\dots,w_k |\nonumber\\
&&  \sigma_2=\sigma_1\circ((n+1)\, z_1 z_2\dots z_k),
\nonumber\\
&&  \tau_2=\tau_1\circ((m+1)\, w_1 w_2\dots w_k\} \rho_{sp}^k
\nonumber\\
&+&
 \sum_{k=0}^\infty \#\{z_1,z_2,\dots,z_{k-1}\;,w_1,w_2,\dots,w_k |\nonumber\\
&&  \sigma_2=\sigma_1\circ((n+1)\, z_1 z_2\dots z_{k-1}),
\nonumber\\
&&  \tau_2=\tau_1\circ((m+1)\, w_1 w_2\dots w_k\} {\rho_{sp}'}^k
\nonumber\\
&+&
 \sum_{k=0}^\infty \#\{z_1,z_2,\dots,z_k\;,w_1,w_2,\dots,w_{k-1} |\nonumber\\
&&  \sigma_2=\sigma_1\circ((n+1)\, z_1 z_2\dots z_k),
\nonumber\\
&&  \tau_2=\tau_1\circ((m+1)\, w_1 w_2\dots w_{k-1}\} {\rho_{sp}'}^k
\label{FMTXMN}
\end{eqnarray}
The value $k=0$ was included in the last two terms.
This is the contribution of the ``simple singular points'' on the
fold.

\section*{Appendix B : Calculation of the eigenvalues of the fold matrix}
We will first calculate the eigenvalues of the fold-matrix in the case $m=0$.
We will start with the non-restricted fold matrix (which is the restricted
fold matrix $F_{(\phi=1,\psi=1)})$.

For $m=0$, the folds always have a common sheet, say $d_1$.
We will assume that at the fold $\gamma_i$ --
$d_1$ is attached to $c_{\alpha_i}$
with $1\leq\alpha_i\leq n+1$ ($i$ being the index of the fold arc).
We will also assume that $a_j$ is joined
to $c_{\sigma_i(j)}$ with
$\sigma_i : \{1,\dots,n\}\rightarrow\{1,\dots,n+1\}$.
{}From the previous chapter
(when we set $\rho_{sp}$ and $\rho_{sp}'$ to zero in (\ref{FMTXMN}, since
there are no spiral points with $k>0$ for $m=0$, and $\tau$
can be dropped as well when $m=0$.)
we see that $\sigma_i$ must satisfy
\begin{eqnarray}
\sigma_i(x) &=& \sigma_{i+1}(x) \mbox{ for
  $x \not\in ( \sigma_i^{-1}(\alpha_i),\sigma_{i+1}^{-1}(\alpha_{i+1}))$}
  \nonumber \\
\sigma_i(\sigma_{i+1}^{-1}(\alpha_{i+1})) &=& \alpha_i \nonumber \\
\sigma_{i+1}(\sigma_i^{-1}(\alpha_i)) &=& \alpha_{i+1} \label{ADJSIG}
\end{eqnarray}

and we get:
\begin{equation}
F^{\sigma_1}_{\sigma_2} =
\sum_{y=1}^n\delta(\sigma_2^{-1}\sigma_1((n+1)y))
=D_{Reg}(\sum_{y=1}^n ((n+1)y)
\label{FMTX0}
\end{equation}
In the sum $\sum_y$ only $y=\sigma_1^{-1}\sigma_2(n+1)$ can give a nonzero
$\delta$-function, and $D_{reg}(\sigma)$ is the matrix of the permutation
$\sigma\in S_{n+1}$ in the regular $(n+1)!\times (n+1)!$ representation.

The regular representation of $S_k$ can be decomposed into irreducible
representations. It is best to use the Yamanouchi formalism \cite{REPRES},
in which case the representations are labeled by the value of the sums
of the permutations in all equivalence classes of $S_k$, and to those are
added the sums of the permutations in all equivalence classes of $S_r$ for
$r=1,\dots,k-1$, thus obtaining a maximal commuting set of operators.
In this representation, for $S_{n+1}$:
\begin{eqnarray}
\sum_{r=1}^n (r,(n+1)) &=& \hlf\sum_{1\leq x,y\leq (n+1)}(xy)
-\hlf\sum_{1\leq x,y\leq n}(xy) \nonumber\\
&=& \lambda_{(2)}^{(v)}-\lambda_{(2)}^{(v^\prime)}\nonumber\\
&=&\hlf +\hlf\sum_{l=1}^q(v_l(v_l-2l)-v_l^\prime(v_l^\prime-2l))\nonumber\\
&=&\hlf+\hlf(v_t(v_t-2l)-(v_t-1)(v_t-1-2l)) = v_t-t \nonumber\\
&&
\end{eqnarray}
Where $v$ is a partition $n+1=\sum_{l=1}^q v_l$ which represents a Young
diagram for $S_{n+1}$ and $v^\prime$ is a Young diagram for $S_n$ that is
obtained from $v$ by taking off one square in an allowed way (that is, leaving
$v_1^\prime\geq v_2^\prime\geq\dots\geq v_l^\prime$) which we denote by
$v_t^\prime=v_t-1$. $\lambda_{(2)}^{(v)}$ is the character for $2$-cycles
in $(v)$, that is the scalar that represents
$\hlf\sum_{1\leq x,y\leq (n+1)}(xy)$ in the $(v)$ representation.
Likewise $\lambda_{(2)}^{(v^\prime)}$ represents
$\hlf\sum_{1\leq x,y\leq n}(xy)$ in the $(v^\prime)$ representation.

We see that the eigenvalues turn out to be $(v_t-t)$, and they come in pairs
of positive-negative eigenvalues. The eigenvalue of $(t-v_t)$ corresponds
to a Young tableau that is the mirror of the first Young tableau.

\subsection*{The Restricted Fold Matrices}
To calculate (\ref{FOLDSP})  we need the eigenvalues of all of the
restricted fold matrices $F_{(\phi\in S_{n+1},\psi\in S_n)}$ (we omit
$\theta,\beta$ since they are trivial for $m=0$).

In the following equations we consider $\psi\in S_n\subset S_{n+1}$.

For given $\phi\in S_{n+1}$ and $\psi\in S_n$ we are interested in the
$\sigma_i$-s that satisfy
\begin{equation}
\sigma_i^{-1}\phi\sigma_i=\psi
\end{equation}
(see (\ref{RESTRF})).

This means that unless $\psi$ and $\phi$ are of the same equivalence class
the restricted fold matrix is identically zero. So suppose that
\begin{equation}
\psi = \rho^{-1}\phi\rho
\end{equation}
so that we are interested only in $\sigma_i=\rho\theta_i$ where
$\theta_i\in S_{n+1}$ is a permutation that commutes with $\psi$.
{}From (\ref{FMTX0}) and (\ref{RESTRF}) we obtain (recall that
$(x,y)$ is the transposition permutation that switches $x$ with $y$):
\begin{eqnarray}
(F_{(\psi,\phi)})^{\theta_1}_{\theta_2} &=&
\delta(\sigma_2^{-1}\circ\sigma_1\circ ((n+1)\sigma_2(n+1)))
   \nonumber\\
&=& \delta(\sigma_2^{-1}\circ(\sigma_1(n+1)\sigma_2(n+1))\circ\sigma_1)
   \nonumber\\
&=& \delta(\theta_2^{-1} \rho^{-1}\circ
(\sigma_1(n+1)\sigma_2(n+1))\circ\rho\theta_1)  \nonumber\\
&=& \delta(\theta_2^{-1}\circ(\theta_1(n+1)\theta_2(n+1))\circ\theta_1)
\nonumber\\
&=& (F_{(1,1)})^{\theta_1}_{\theta_2}
\label{RESTHTH}
\end{eqnarray}
We have changed the indices of $F$ to $\theta_1,\theta_2$ since the
$\sigma_i$ that come into consideration are in one to one correspondence
with the $\theta_i$-s.

This means that $F$ can be put in block form as follows :

The equivalence class of $\psi$ is in one to one correspondence with the
partitions of $n$ since
\begin{equation}
\psi = (a_1)(a_2)\dots(a_{r_1}) (b_1c_1)(b_2c_2)\dots(b_{s_1}c_{t_1})
\dots (3-cycles) \dots (4-cycles) \dots
\end{equation}
such that $(n+1)$ is included in the $1-cycles$.
The permutations $\tau$ that commute with $\psi$ are
those that transform the indices of the cycles into other indices that
describe the same cycle. This means, in particular, that if
$(a_{i1},a_{i2},\dots,a_{ik})$ are all the $k-cycles$ with $i=1,\dots,N_k$,
then $\tau(a_{ij})=a_{i^\prime j^\prime}$ for some $i^\prime j^\prime$ and
if $\tau(a_{ij})=a_{i^\prime j^\prime}$ then
$\tau(a_{ir})=a_{i^\prime (j^\prime+r-j \pmod{k})}$ so as to transform the
$i$-th $k-cycle$ to the $i^\prime$-th $k-cycle$.

What pairs of $\theta_1,\theta_2$ that commute with $\psi$ give a non-zero
restricted fold matrix element?
According to (\ref{RESTHTH}) $\theta_1$ and $\theta_2$ must differ just
by switching two elements which are $\theta_1(n+1)$ and $\theta_2(n+1)$.
But, since they commute with $\psi$, $\theta_1(n+1)$ and $\theta_2(n+1)$ must
be $1$-cycles of $\psi$.
Let $N_k(\psi)$ be the number of $k$-cycles of $\psi$.
Thus, the restricted fold matrix is equal to
$\prod_{k=2}^n (N_k(\psi)^k N_k(\psi)!)$ copies
of the (non-restricted) $N_1(\psi)!\times N_1(\psi)!$ fold matrix.
This is because each pair $\theta_1,\theta_2$ that gives a nonzero
matrix element in (\ref{RESTHTH}) can be written as
\begin{eqnarray}
\theta_1 &=& \theta_1'\circ\theta_1'' \\
\theta_2 &=& \theta_2'\circ\theta_2''
\end{eqnarray}
where $\theta_1',\theta_2'$ are permutations of the $N_1(\psi)$ $1$-cycles
of $\psi$ and $\theta_1'',\theta_2''$ are permutations among the $k>1$ cycles.
according to what was said above $\theta_1''=\theta_2''$ in  order to give
a non-zero matrix element.
The number $\prod_{k=2}^n (N_k(\psi)^k N_k(\psi)!)$ is the number
of possible values of $\theta_1''$.

\subsection*{A General $(n,m)$ Region}
If we do not allow ``spiral'' singular points, that is, put
$\rho_{sp}=0$ in (\ref{FMTXMN}), we get a fold matrix that at each singular
point leaves one of the covers that joins a fold unchanged.

That is, the matrix elements $F^{\sigma_1\tau_1}_{\sigma_2\tau_2}$
are sums of two terms, one is proportional to $\delta(\tau_1^{-1}\tau_2)$,
and the other is proportional to $\delta(\sigma_1^{-1}\sigma_2)$.
In this case we can write $F(n,m)$ as a sum of tensor products:
\begin{equation}
F(n,m) = F(n,0) \otimes 1 + 1\otimes F(0,m)
\end{equation}
where $1$ is the identity matrix. The eigenvalues of $F(n,m)$ are the
sums of eigenvalues of $F(n,0)$ and $F(0,m)$.

\end{document}